\def\be{\begin{equation}}
\def\te{\end{equation}}
\def\bea{\begin{eqnarray}}

\def\tea{\end{eqnarray}}

\documentclass[
superscriptaddress,
preprint,
nofootinbib,
amsmath,amssymb,
aps,
prd,
floatfix,
]{revtex4-2}
\usepackage{graphicx}
\usepackage{subfigure,upgreek,enumerate,color,verbatim,comment,bigints,stackrel}
\usepackage{bm,pbox,cancel}
\usepackage[dvipsnames]{xcolor}
\usepackage[colorlinks=true,hyperfootnotes=true,breaklinks=true,citecolor=blue,linkcolor=PineGreen]{hyperref}
\usepackage{physics}

\begin{document}
\count\footins = 1000
\newcommand{\tbf}[1]{\textbf{#1}}
\title {Heat Capacity and Quantum Compressibility of \\ Dynamical Spacetimes with Thermal Particle Creation}
\author{Jen-Tsung Hsiang}
\email{cosmology@gmail.com}
\affiliation{College of Electrical Engineering and Computer Science, National Taiwan University of Science and Technology, Taipei City, Taiwan 106, ROC}

\author{Yu-Cun Xie}
\email{xyc@terpmail.umd.edu}
\affiliation{Department of Physics, University of Maryland, College Park, Maryland 20742, USA}

\author{Bei-Lok Hu}
\email{blhu@umd.edu}
\affiliation{Maryland Center for Fundamental Physics and Joint Quantum Institute, University of Maryland, College Park, Maryland 20742, USA}

\date{v1 Sept 3, 2023. v2 Jan 24, v3 March 29, v4 April 24, 2024} 

\begin{abstract} 
This work continues the investigation in two recent papers on the quantum thermodynamics of spacetimes, 1) placing what was studied in \cite{CHH1} for thermal quantum fields in the context of early universe cosmology, and  2) extending the considerations of vacuum compressibility of  dynamical spaces treated in \cite{XHH1}  to dynamical spacetimes with thermal quantum fields. We begin with a warning that thermal equilibrium condition is not guaranteed to exist or maintained in a dynamical setting and thus finite temperature quantum field theory in cosmological spacetimes needs more careful considerations than what is often described in textbooks. A full description requires nonequilibrium quantum field theory in dynamical spacetimes using `in-in' techniques. A more manageable subclass of dynamics is where thermal equilibrium conditions are established at both the beginning and the end of evolution, where the in-thermal state and the out-thermal state are both well defined. Particle creation in the full history can then be calculated in this in-out  asymptotically-stationary setup  via the $S$-matrix transition amplitudes.  Here we shall assume an in-vacuum state.  It has been shown that if the intervening dynamics has an initial period of exponential expansion, such as in inflationary cosmology, particles created from the parametric amplification of the vacuum fluctuations in the initial vacuum will have a thermal spectrum measured at the out-state. Under these conditions finite temperature field theory can be applied to calculate the quantum thermodynamic quantities. Here we consider a massive conformal scalar field in a closed four-dimensional Friedmann-Lema\^itre-Robertson-Walker universe based on the simple analytically solvable Bernard-Duncan model. We calculate the energy density of particles created from an in-vacuum and derive the partition function. From the free energy we then derive the heat capacity and the quantum compressibility of these spacetimes with thermal particle creation. We end with some discussions and suggestions for further work in this program of studies.  
\end{abstract}

\maketitle
\parskip=10pt

\newpage
\tableofcontents
\newpage

\section{Introduction}

This is a sequel to two recent papers on the quantum thermodynamics of spacetimes,  1)  placing the work of \cite{CHH1} for the heat capacity and quantum compressibility of {static} spaces containing quantum fields at finite temperatures in the context of early universe cosmology, and 2) extending the dynamic compressibility of spaces containing quantum fields in a vacuum state considered in \cite{XHH1} to thermal quantum fields.  We have considered in \cite{XHH1} three types of early universe vacuum quantum processes:  the Casimir effect, the trace anomaly \cite{FHH79,And83,And84} and vacuum particle creation \cite{ZelSta71,HuPar77,HuPar78,HarHu79,HarHu80,And85} in a number of dynamical spaces.  Here, we shall focus only on thermal particle creation in a Friedmann-Lema\^itre-Robertson-Walker (FLRW) universe, with special attention paid to inflationary dynamics with a period of  exponential expansion.   

Unlike the settings in the previous two papers, either thermal quantum fields in a static spacetime or vacuum quantum field processes in a dynamical spacetime, finite temperature is predicated upon the existence of a thermal equilibrium condition,  which cannot be assumed a priori for any dynamical spacetime. Even when such a condition is fulfilled at one time, particle creation at that moment of time will steer the system away from equilibrium in a later moment.  Thus before one discusses the thermodynamic properties of the universe one needs to first address this basic conceptual issue of when in a dynamical spacetime could thermal equilibrium  be conditionally established, and under what conditions can a thermal equilibrium state be maintained throughout its  evolution.

\subsection{Dynamics and Thermal Field Theory}

Thermal field theory deals with systems at finite temperatures. This is predicated upon an underlying assumption  that the system can be kept in thermal equilibrium with a finite temperature bath. For time-dependent systems involved in cosmological particle creation (CPC) and dynamical Casimir effect (DCE), one key issue which need be addressed ab initio is: how can one ensure cosmological particle creation or enforce dynamical Casimir effect such that the equilibrium condition is maintained in the temporal evolution of the system. This is possible if there is no energy input into, or entropy generated from, the evolving system in the process. Otherwise thermal field theory is inadequate and one needs to use concepts in nonequilibrium (NEq) statistical mechanics and techniques in nonequilibrium quantum field theory. In the cosmology setting, this issue has been addressed in a series of papers \cite{Hu81,Hu82A,Hu82B,Hu83} by one of the present authors. We summarize the key points from that body of work relevant to our investigation here: 

\subsubsection{Finite temperature quantum field theory in cosmology}

In an evolving universe, allowing for particle creation from the vacuum, the number of particles, the energy and entropy of quantum matter may all change. If the expansion (or contraction) is isotropic such as in a FLRW universe, and if the particles described by a free field obey conformally-invariant field equations (e.g., photons), it can be shown that no entropy is generated and the gas remains in thermal equilibrium throughout the evolution  \footnote{This is the situation where there is no particle creation. Mathematically, the conformal flatness of the FLRW metric and the conformal invariance of the matter field allow one to perform a conformal transformation on the field and  on the spacetime, giving rise to a field equation in a flat space where there is no particle creation.}.

In the radiation-dominated era  one often encounters the statement that  the (Tolman) temperature $T \propto 1/a$, where $a(t)$ is the scale factor and $t$ the cosmic time. This assumes an isentropic expansion of the universe, where the total number of photons (assuming that is the dominant particle species present) measured by $\Theta \equiv T(t) a(t)$ remains a constant.

For interacting particles, thermal equilibrium can be maintained at a temperature $T(t)$ provided that the characteristic rate of thermal interaction $R$ is much greater than the expansion rate of the universe $H =  (da/dt)/a$,  i.e., $R \gg H$. We shall introduce the criterion of nonadiabaticity and the notion of quasi-equilibrium in this context in the next subsection. 

\subsubsection{Casimir, trace anomaly and particle creation: from vacuum to radiation dominated}

 As a simple illustration of the inter-knitted relation between vacuum effects due to the Casimir energy, the trace anomaly and particle creation, such as studied in \cite{XBH,XHH1}, and thermal effects, we mention the following relation \cite{Hu82B}. Consider a massless conformal scalar field at temperature $T$  in a closed FLRW universe, the total  energy density $\rho$ has vacuum contributions $\rho_0$ comprising the Casimir (CA) energy density and  the trace anomaly (TA) plus contributions from thermal fields $\rho_{\textsc{t}}$  
\begin{equation}
\rho = \rho_0 + \rho_{\textsc{t}} = (\rho_{\textsc{ca}} + \rho_{\textsc{ta}}) + (\rho_{\textsc{r}} + \rho_{\textsc{c}} )\,.
\end{equation} 
where $\rho_{\textsc{r}} = \pi^2 T^4/30$ is the Planck thermal energy density, and $\rho_{\textsc{c}}$ is a correction term which takes on the forms given below at high and low temperatures. The controlling parameter is the photon number $\Theta = T a$, which is a constant for isentropic expansion.

i) In the high-temperature $T$ (or small $a$) limit, large $\Theta$: 
$\rho_{\textsc{c}} \to -\rho_{\textsc{ca}}$ (minus the Casimir energy), and thus $\rho \to  \rho_{\textsc{ta}} + \rho_{\textsc{r}} $.
 
ii) In the low-temperature $T$ (or large  $a$) limit, small $\Theta$:
$ \rho_{\textsc{c}} \to  -\rho_{\textsc{r}}$ (minus the Planck energy) and thus $\rho \to \rho_{\textsc{ta}} + \rho_{\textsc{ca}}$. 

We see the former case corresponds to a universe containing hot radiating gas while the latter corresponds to a cold quantum universe. Any finite value of $\Theta$ corresponds to a universe with some radiation content. The finite-temperature theory of quantum fields thus provides a natural framework for the discussion of the interplay of quantum versus classical processes with vacuum or matter contents, as well as the occurrence and transition from the vacuum-dominated to the  radiation-dominated
regimes of the early universe.

\subsection{Levels of approximation in treating a basically nonequilibrium problem} 
 
In evolutionary cosmology when  the temperature of the universe at a certain time in history  is mentioned one  has already made the assumption that matter at that time has come to thermal equilibrium, which needs justification. Cosmological particle creation  is fundamentally a nonequilibrium quantum field process. 

For non-conformally invariant quantum fields, be it with mass,  or massless but minimally-coupled (e.g., the two polarizations of gravitons can thus be described), or for spacetimes which are not conformally-static (e.g., the Kasner or mixmaster universes), particle production can happen which generates entropy and disrupts the thermal equilibrium condition. A more accurate and thorough treatment would be  by way of nonequilibrium quantum field theory~\cite{CalHu2008}, via the `in-in' formalism as an initial value problem, as mentioned previously.

Having enunciated the conceptual and technical challenges for treating this intrinsically nonequilibrium problem, we now ask, are there situations which allows for a $S$-matrix or `in-out' treatment via finite temperature quantum field theory?   A subclass of dynamics, albeit somewhat artificial, is known as the asymptotically-stationary or statically-bounded setup. It has the distinct advantage that the quantum field theory in both asymptotic regions, thanks to their endowed global timelike Killing vector, are well defined.  Here, one assumes that the in-state and the out-state of the system are in thermal equilibrium at temperatures $T_i$, $T_f$. In the intervening time there is particle production, which  can be calculated from the transition amplitudes of a $S$-matrix.  For thermal scalar fields, the in-state is an $n$-particle state obeying the Bose-Einstein distribution. Particle creation from the vacuum state is known as spontaneous production while from  an $n$-particle state is known as stimulated production. (See, e.g., \cite{HKM94}). The assumption that the out-state is also a thermal state implies that the particles created either spontaneously or by stimulation have had enough time and channels to come to equilibrium.  The temperature of the out-state should be calculable as it is related to the number of particles created, assuming that is the only entropy generating process. A simple-minded way to see this is by way of the Stefan-Boltzmann law, where the energy density is proportional to $T^4$ for a photon gas. In fact, we can track the increase in energy  associated with the number of particles created in real time, from the in- to the out-state, even though temperature is not a well-defined notion in a time-dependent setting. 

\section{Thermal particle creation in inflationary universe}

In the  asymptotically-stationary set up, there is a special class of evolutionary cosmology where the particles created from the in-vacuum  has a thermal spectrum measured in the out state, namely, when there is a period of exponential expansion. Since this includes the widely popular inflationary universe we shall in this paper focus on this class of cosmology for studying the thermodynamic properties of spacetimes due to quantum field processes. Thermal particle creation was shown in the 70s in several statically-bounded models which  are completely solvable so one can closely examine all the details (see, e.g., \cite{KHMR}): One model worked out by Bernard and Duncan \cite{BerDun} is for a massive conformal scalar field in a two-dimensional spatially-flat FLRW universe with a scale factor $a(\eta)$ (conformal time $dt=a\, d\eta$) following a hyperbolic-tangent (conformal) time dependence,
\be\label{BerDun}
    a^2(\eta) = A + B \tanh \rho \eta\,,  
\te
which tends to  constant values $a^2_{\pm} \equiv A \pm B$ at asymptotic times
$\eta \rightarrow \pm \infty$. We see that from an in-static spacetime with scale factor $a_-$  there is an initial  exponential rise measured by the rise parameter $\rho$,  followed by a nearly linear mid-segment and a smooth exit to an out-static spacetime with scale factor $a_+$.  These authors found that if  $a_+ \gg a_-$  to a good approximation the ratio of the modulus of the Bogoliubov coefficients 
\be 
|\beta_k /\alpha_k|^2  = \exp (-2 \pi \omega_{\textsc{in}} /\rho)\,. ~~ 
\te
where $\omega_{\textsc{in}}$ is the mode frequency of the field of the in-state at $\eta\to-\infty$. For high momentum modes, one can recognize the Planckian distribution with temperature given by
\be
k_B T_{\eta} = \frac{\rho}{2 \pi a_+}\,.       
\te
The Bogoliubov coefficients $\alpha_k$, $\beta_k$ are defined as follows:
Call  $\Phi_k^{\textsc{in},\,\textsc{out}}(t, x)$ the positive frequency mode function of the $k^{\text{th}}$ mode of the quantum field in the in- and out-regions at $t= - \infty$ and $t = + \infty$, respectively. They are related by the Bogoliubov coefficients $\alpha_k$, $\beta_k$ defined by
\be
\Phi_k^{\textsc{in}} (t, x) = \alpha_k \Phi_k^{\textsc{out}} (t,x) + \beta_k \Phi_{-k}^{\textsc{out} *}
(t,x)\,.
\te 
The probability $P_n(\vec k)$ of observing $n$ particles in mode $\vec k$ at late times is \cite{Par76}:
\be
P_n(\vec k) = |\beta_k /\alpha_k|^{2n} |\alpha_k|^{-2}.
\te
{}From this one can find the average number of particles $\langle N_{\vec k}
\rangle$ created in mode $\vec k$ (in a comoving volume) at late times to be
\be
N_{\vec k}\rangle = \sum_{n=0}^\infty n\,P_n (\vec k) = 
|\beta_k|^2 \,.
\te
The modulus of their ratio is particularly useful for identifying the thermal character of particle creation. 

In another analytically solvable model, Parker \cite{Par77} treated a massless minimally coupled scalar field in a FLRW universe using a  more complicated functional form  for the scale factor's $\tau$ time defined by $d\tau=a^3dt$ dependence.  He showed that significant particle creation occurs during the early period when the universe expands exponentially fast. In fact  the  ratio of the modulus of the Bogoliubov coefficients obtained in a simpler model where the scale factor takes on an initial exponential rise  has the same form as in this more complicated model (see also Berger \cite{Berger}). Both cases studied give rise to a thermal spectrum with temperatures proportional to the rise function, e.g., $\rho$ in conformal time $\eta$ in the Bernard-Duncan model, and $\sigma$ in $\tau$ time in the Parker-Berger model.   

\subsection{General properties of cosmological scale function for thermal production}

In more general terms,  Parker speculated that the exponential form in $|\beta /\alpha|$
should hold for a general class of scale functions which possess the properties that: 1) they
smoothly approach a constant at early time, 2) their values at late times
are much larger than at initial times, and 3) they and their derivatives are continuous functions.
The exponential factor contained in the scale functions at early times is  responsible
for the thermal property of particle creation, with the temperature
proportional to the rise factor in the exponential function. 
More noteworthy is that  this property is quite insensitive to the late time asymptotically
static behavior of $a (\tau)$, namely, it could be the asymptotically flat behavior of a $\tanh$ function, or the continuing rising behavior of an $\exp$ function.
     
\subsection{Heat capacity and quantum compressibility}

We have specified the setting and identified the focus of our problem, namely, calculate the energy density of quantum field processes ranging from the Casimir effect to the trace anomaly and particle creation.  We have chosen to work with the class of dynamics where there is a period of exponential expansion which gives rise to a thermal spectrum of particle production. This thermality condition facilitates a simpler calculation of the heat capacity and the quantum compressibility of this class of cosmology. Focusing only on particle production we can break this down into several steps:

1) For the in-out equilibrium state set up, assuming we begin with the vacuum, i.e.,  temperature $T_{\textsc{in}}=0 $, calculate the energy density of particles created from this in-vacuum. For the class of dynamics with a period of exponential expansion, call it inflationary cosmology,  the particles produced has a thermal spectrum, the temperature of which is determined by the expansion rate. Even for a dynamics which does not become asymptotically-stationary at late times, per what was said above, an initial exponential expansion period would produce particles with a thermal spectrum the temperature of which we can use for our quantum thermodynamics investigations.

2 ) Heat capacity is defined under two separate conditions, constant volume and constant pressure. For static spacetimes these are well defined, see, e.g., \cite{CHH1}. For  spaces whose volumes change with time, we have to make sure that all cases in the exponential expansion family (we may call this inflationary universe) with different rise parameters expand to the same volume in the out-state. Integrating the energy density over the same volume gives the energies in different inflationary universes with varying expansion rates. The faster the expansion the larger amount of particles produced, the bigger the energy and the higher the temperature. The heat capacity of an inflationary universe is obtained by calculating the change of the energy from one member of the family to another with respect to the temperature change across those two relevant members of the family.  

3) To calculate the thermal compressibility, one needs the Helmholtz free energy. From the particle number one can calculate the entropy\footnote{In a textbook description it is often stated that the entropy of a photon gas is given by the number of photons. Note this assumes many layers of coarse graining from its point of origin, namely, the vacuum. Entropy of particles created from the vacuum is a conceptually subtle issue. Since the vacuum is a pure state particle pairs created from it remains so -- there should be no entropy generation in vacuum particle production. Only after some coarse-graining measure, such as paying attention only to the particle number but ignoring the phase coherence relation or correlations between the particles \cite{HuPav86,HuKan87,KME98} or if only one particle in the pair is observable,  would such an observer report on entropy being generated \cite{CamPar,LCHparent}. For a summary description of this subject, see \cite{HHentropy}.} Combining the two one obtains the Helmholtz free energy and from there one can calculate all the thermodynamic quantities.

In the next section we proceed to analyze thermal particle creation of a massive conformal scalar field in a four-dimensional closed Friedmann-Lema\^itre-Robertson-Walker universe based on the simple analytically solvable Bernard-Duncan model which is statically-bounded with an initial period of exponential expansion. We calculate the energy density of particles created from the in-vacuum and derive the partition function. From the Helmholtz free energy we then derive the heat capacity and the quantum compressibility of these spacetimes with thermal particle creation.  In the last section we conclude with  some discussions and suggestions for further work in this program of studies.         

\section{Quantum thermodynamics of an inflationary universe }

Consider a  massive ($m$) scalar field $\phi$ in an expanding four-dimensional closed Friedmann-Lema\^itre-Robertson-Walker (FLRW) universe with line element \begin{equation}
    ds^2= -dt^2 + a^2(t)\,d\Omega^2_{S^3} =  a^2(\eta)\,(-d\eta^2+d\Omega^2_{S^3})\,,
\end{equation}
where  $d\Omega^2_{S^3}$ is the line element for a 3-sphere $S^3$, $t$ is the cosmic time and  $\eta \equiv\displaystyle \int\!dt/a(t) $ is the conformal time, This field obeys the wave equation:
\begin{align}\label{E:grit}
    \frac{1}{\sqrt{-g}}\,\partial_{\mu}\biggl[\sqrt{-g}\,g^{\mu\nu}\partial_{\nu}\phi\biggr]-m^2\phi-\xi R\,\phi=0\,,
\end{align}      
where $\sqrt{-g}=a^4$,  $g^{\mu\nu}=a^{-2}\mathsf{g}^{\mu\nu}$ and $\mathsf{g}^{\mu\nu}$ is the metric corresponding to the line element $-d\eta^2+d\Omega^2_{S^3}$. The Ricci scalar  $R$ is given by
\begin{equation}
    R=\frac{6}{a^2}\biggl(1+\frac{a''}{a}\biggr)\,,
\end{equation}
where a prime denotes taking the derivative with respect to the conformal time $\eta$. We shall consider conformally-coupled fields where  $\xi=1/6$ in four dimension.

Spelling out \eqref{E:grit} explicitly gives
\begin{align}
    -\frac{1}{a^4}\,\partial_\eta\Bigl(a^2\,\partial_{\eta}\phi\Bigr)+\frac{1}{a^2}\bm{\nabla}_{S^3}^2\phi-m^2\phi-\frac{1}{a^2}\Bigl(1+\frac{a''}{a}\Bigr)\,\phi&=0\,,
\end{align}   
where 
\begin{equation}
    \bm{\nabla}_{S^3}^2=\partial_i\bigl(\mathsf{g}^{ij}\partial_j\bigr)\,.
\end{equation}
Now let $\phi=\chi/a$, and Eq.~\eqref{E:grit} becomes
\begin{align}
    \chi''-\bm{\nabla}_{S^3}^2\chi+\bigl(m^2a^2+1\bigr)\chi&=0\,.\label{E:dieur}
\end{align}
Since the eigenvalues $\lambda_{l}$ corresponding to $\bm{\nabla}_{S^D}^2$ are $\lambda_{l}=-l(l+n-1)$ for $l=0$, 1, 2, $\cdots$. In the case $D=3$, Eq.~\eqref{E:dieur} reduces to
\begin{align}
    \chi''(\eta)+\omega_l^2\,\chi(\eta)&=0\,,&\omega_l^2=(l+1)^2+m^2a^2\,.
\end{align}
The corresponding physical frequency is $\varpi_l=\dfrac{\omega_l}{a}=\sqrt{\dfrac{(l+1)^2}{a^2}+m^2}$.

\subsection{Asymptotically-stationary universe with a period of exponential expansion}

As an example of an asymptotically-stationary universe, we let the scale factor  $a(\eta)$ evolve in conformal time as
\begin{equation}
    a^2(\eta)=\frac{a_f^2+a_i^2}{2}+\frac{a_f^2-a_i^2}{2}\,\tanh\frac{\eta}{\Delta}\,,
\end{equation}
where the scale factor transits from one constant value $a_i$ in the asymptotic past to another constant $a_f$ in the asymptotic future. In the transition time of duration $\Delta$ (inverse of the rise function), the universe expands exponentially fast, enabling us to explore the thermodynamic properties. Note, however, the exponential expansion considered here is in conformal time, not in cosmic time. (For the standard inflationary cosmology $a(t)=e^{Ht}=-1/(H\eta)=a(\eta)$ so the exponential form is not reserved.)

Since the equation of motion for $\chi$ is exactly the same as in the Bernard-Duncan model \cite{BerDun} in a two-dimensional, spatially-flat FLRW spacetime. With the continuous $k$ there replaced by the discrete values $k_l=l+1$ here, their results apply here. Let
\begin{align}\label{E:potye}    \omega_l^{\textsc{in}}&=\sqrt{k_l^2+m^2a_i^2}\,,&\omega_l^{\textsc{out}}&=\sqrt{k_l^2+m^2a_f^2}\,,&\omega_l^{\pm}&=\frac{\omega_l^{\textsc{out}}\pm\omega_l^{\textsc{in}}}{2}\,,
\end{align}
the Bogoliubov coefficients $\alpha_{k_l}$, $\beta_{k_l}$ in the asymptotic future are given by
\begin{align}\label{E:bdire}
    \lvert\alpha_{k_l}\rvert^2&=\frac{\sinh^2(\pi\omega_l^+\Delta)}{\sinh(\pi\omega_l^{\textsc{in}}\Delta)\,\sinh(\pi\omega_l^{\textsc{out}}\Delta)}\,,&\lvert\beta_{k_l}\rvert^2&=\frac{\sinh^2(\pi\omega_l^-\Delta)}{\sinh(\pi\omega_l^{\textsc{in}}\Delta)\,\sinh(\pi\omega_l^{\textsc{out}}\Delta)}\,,
\end{align}
with the Wronskian condition, $\lvert\alpha_{k_l}\rvert^2-\lvert\beta_{k_l}\rvert^2=1$. This follows from the Bogoliubov transformation
\begin{equation}
    u_{k_l}^{\textsc{in}}(\eta,\bm{x})=\alpha_{k_l}\, u_{k_l}^{\textsc{out}}(\eta,\bm{x})+\beta_{k_l}\, u_{-k_l}^{\textsc{out}*}(\eta,\bm{x})\,,
\end{equation}
where the positive frequency mode $ u_{k}(\eta,\bm{x}) \propto e^{-i\omega_l\eta}$ in both asymptotic regions for the suitable $\omega_l$. The particle number generated during the transition is given by $\lvert\beta_{k_l}\rvert^2$ for mode $k_l$. 

The energy associated with particle creation
is obtained by summing the contributions from all modes
\begin{equation}
    E=\sum_{l=0}^{\infty}d_l^{(4)}\Bigl(\lvert\beta_{k_l}\rvert^2+\frac{1}{2}\Bigr)\,\varpi_l
\end{equation}
where the degeneracy $d_l^{(D+1)}$ in $R\times S^{D}$ is
\begin{equation}
	d_{l}^{(D+1)}=(2l+D-1)\,\frac{(l+D-2)!}{l!(D-1)!}=(l+1)^2\,,
\end{equation}
for $D=3$ and $l=0$, 1, 2, $\cdots$. This quantity associated with vacuum particle creation is the source of our present investigation into the quantum thermodynamics of dynamical spacetimes. Note, however,  this expression   does not include other contributions to the vacuum energy such as the trace anomaly. {In fact, there is no contribution from the trace anomaly in the asymptotically stationary out-state because it only depends on different orders of time derivatives of the scale factor.}

\subsection{Thermal particle production originating from the in-vacuum}

Let us try to understand the physics of particle creation associated with a universe undergoing exponential expansion by examining $\lvert\beta_{k_l}\rvert^2$ given by Eq.~\eqref{E:bdire}. In the case of a large $\Delta$, corresponding to a very long duration of slow transition or gradual rise:
\begin{align}\label{E:rrrhr}
    \pi\omega_l^{\textsc{out}}\Delta&\gg1\,,
\end{align}
we can approximate $\lvert\beta_{k_l}\rvert^2$ by
\begin{align}
    N_l=\lvert\beta_{k_l}\rvert^2\sim\frac{1}{2}\,\coth(\pi\omega_l^{\textsc{in}}\Delta)-\frac{1}{2}\,.
\end{align}   
We recognize this as the distribution of the mean particle number of bosons at finite temperatures. Thus at late times, in a sufficiently high frequencies {in the out-state}, we may identify an {\it effective temperature}  from the statistical distribution of the created particles in the out state
\begin{equation}\label{E:evsdsd}
    \frac{\beta_{\textrm{eff}}\,\varpi_l}{2}=\pi\omega_l^{\textsc{in}}\Delta\,,
\end{equation}
given by
\begin{equation}
    T_{\textrm{eff}}=\frac{1}{2\pi a_f\,\Delta}\frac{\omega_l^{\textsc{out}}}{\omega_l^{\textsc{in}}}\,.
\end{equation}
{In particular, for lower $l$ modes}, $l\ll m\,a_i$, the effective temperature $T_{\textrm{eff}} \sim\dfrac{1}{2\pi a_i\,\Delta}$  becomes independent of field parameters.

As a check, that there is no particle creation for a massless field, notice Eq.~\eqref{E:potye} gives $\omega_l^{\textsc{in}}=\omega_l^{\textsc{out}}$ for all values of $l$, so we have $\omega_l^-=0$ and  $\lvert\beta_l\rvert^2=0$. This is a well-known fact, that there is no particle creation for a conformally-coupled massless field in a conformally-static spacetime.

In the other extreme, for a very small $\Delta$, that is, a remarkably rapid transition or a very steep rise, with 
\begin{align}\label{E:riyhr}
    \pi\omega_l^{\textsc{out}}\Delta&\ll1\,,
\end{align}
we find
\begin{equation}\label{E:eotgd}
     N_l=\lvert\beta_{k_l}\rvert^2\sim\frac{(\omega_l^{\textsc{out}}-\omega_l^{\textsc{in}})^2}{4\omega_l^{\textsc{out}}\omega_l^{\textsc{in}}}\,.
\end{equation}
This is a rather unusual situation because it yields a constant independent of $\Delta$, hence independent of the effective temperature. However, with this fairly strong constraint, it automatically implies $\pi\omega_l^{\textsc{in}}\Delta\ll1$. If we write Eq.~\eqref{E:riyhr} into
\begin{align}
    \frac{1}{\pi\omega_l^{\textsc{out}}\Delta}=\biggl(\frac{1}{2\pi a_f\,\Delta}\frac{\omega_l^{\textsc{out}}}{2\omega_l^{\textsc{in}}}\biggr)\frac{\omega_l^{\textsc{in}}a_f}{\omega_l^{\textsc{out}}{}^2}\gg1\,,
\end{align}
and note that the expression inside the parentheses may be identified as the effective temperature $T_{\textrm{eff}}$ of the field at late time after the transition stops, then Eq.~\eqref{E:riyhr} is equivalent to requiring
\begin{equation}
    T_{\textrm{eff}}\,\frac{\omega_l^{\textsc{in}}a_f}{\omega_l^{\textsc{out}}{}^2}\gg1.
\end{equation}
The second factor is typically very tiny on considering the extreme expansion of the spacetime. This then infers that the  effective temperature of the environment is exceptionally high, which may not be a sensible scenario for the theory under consideration. Alternatively we can imagine that the spacetime in this limit needs to expand with a very large scale change over an exceeding short time in a way the scale factor behaves like a step function in time. Thus, it is unlikely that the field can still be assumed to be globally spatially homogeneous after transition, and expectantly large field fluctuations may be induced. They can persist for a long time if there is no interacting environment to relax them.

\begin{figure}
	\centering
	\includegraphics[width=0.65\textwidth]{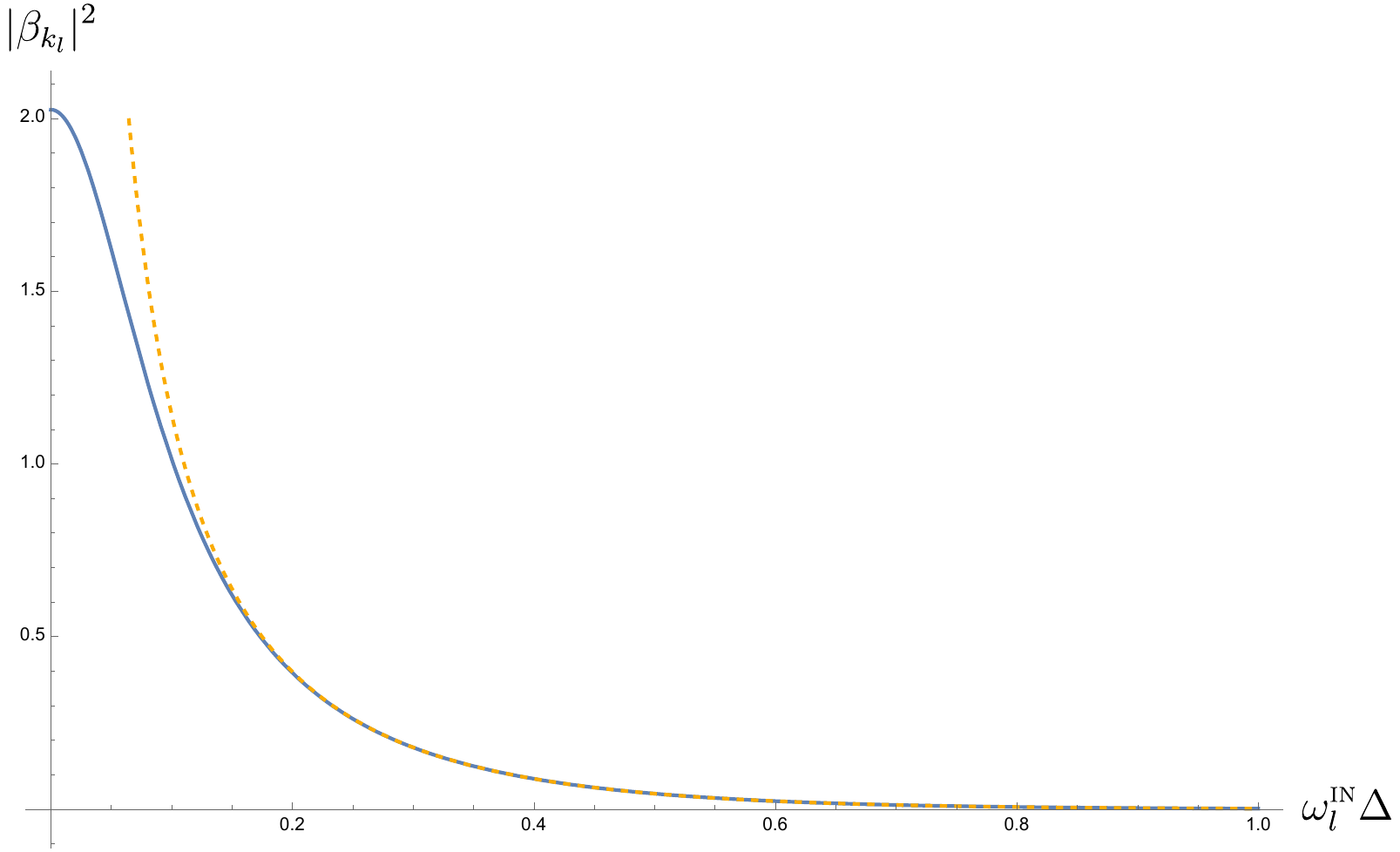}
	\caption{Dependence of $\lvert\beta_{k_l}\rvert^2$ on $\Delta$ is shown in the blue curve. For comparison, the corresponding result according to the Bose-Einstein distribution is shown in the orange dashed curve. They deviate at very small $\Delta$, given by Eq.~\eqref{E:eotgd}. Here we choose $\omega_l^{\textsc{out}}=10$ and $\omega_l^{\textsc{in}}=1$.}\label{Fi:betal}
\end{figure}

Instead we may ease up the limit a bit by demanding $\pi\omega_l^{\textsc{in}}\Delta\ll1$ only, and then we obtain
\begin{equation}\label{E:bksfser}
    N_l=\lvert\beta_{k_l}\rvert^2\sim\tanh\frac{\pi\omega_l^{\textsc{out}}\Delta}{2}\,\frac{1}{2\pi\omega_l^{\textsc{in}}\Delta}\,.
\end{equation}
The second factor is nothing but
\begin{equation}
    \frac{1}{2\pi\omega_l^{\textsc{in}}\Delta}=\frac{\omega_l^{\textsc{out}}}{2\pi a_f\Delta\omega_l^{\textsc{in}}}\frac{a_f}{\omega_l^{\textsc{out}}}=\frac{T_{\textrm{eff}}}{\varpi_l}\,,
\end{equation}
Here $\varpi_l=\omega_l^{\textsc{out}}/a_f$ is the physical frequency of mode $l$ at late times after the transition ends. This is the expression we typically see for the particle number in the high temperature limit. Thus, if Eq.~\eqref{E:bksfser} is to describe the well known high temperature limit, we should require
\begin{equation}
    \tanh\frac{\pi\omega_l^{\textsc{out}}\Delta}{2}\simeq1\,.
\end{equation}
In other words, by the high temperature limit in the conventional sense, we mean $\pi\omega_l^{\textsc{in}}\Delta\ll1$ as well as $\pi\omega_l^{\textsc{out}}\Delta\gg1$. Thus, in summary, under the assumption $\pi\omega_l^{\textsc{out}}\Delta\gg1$, the various temperature limits are determined by $\pi\omega_l^{\textsc{in}}\Delta$: The low temperature regime corresponds to $\pi\omega_l^{\textsc{in}}\Delta\gg1$ while the high temperature regime occurs when $\pi\omega_l^{\textsc{in}}\Delta\ll1$.

\subsection{Energy, pressure, entropy, heat capacity, quantum compressibility}

Having established the linkage in Eq.~\eqref{E:evsdsd} with thermal field theory, we may now explore the quantum thermodynamics of exponentially expanding spacetimes with thermal particle creation.  Since we are working with the out-state where spacetime is stationary the partition function would be similar to that in Minkowski space thermal field theory: 
\begin{equation}
    \mathcal{Z}=\prod_{l=0}^{\infty}\biggl[\sum_{N_l=0}^{\infty}e^{-(N_l+1/2)\beta_{\textrm{eff}}\varpi_l}\biggr]^{d_l^{(4)}}=\prod_{l=0}^{\infty}\biggl[\frac{e^{\frac{\beta_{\textrm{eff}}\varpi_l}{2}}}{e^{\beta_{\textrm{eff}}\varpi_l}-1}\biggr]^{d_l^{(4)}}\,.
\end{equation}
The summation is over all modes even though the high frequency modes are less easily excited while their contributions are suppressed by a Boltzmann-like factor. The Helmholtz free energy $\mathcal{F}$ associated with thermal particle production then follows
\begin{align}
    \mathcal{F}&=-\frac{1}{\beta_{\textrm{eff}}}\,\ln\mathcal{Z}=\sum_{l=0}^{\infty}d_l^{(4)}\biggl[\frac{\varpi_l}{2}+\frac{1}{\beta_{\textrm{eff}}}\ln\bigl(1-e^{-\beta_{\textrm{eff}}\varpi_{l}}\bigr)\biggr]\,.
\end{align} 
The first term in the square brackets gives the vacuum contribution from the zero-point energy of each mode at the out-state. The finite-temperature contribution is from the second term. Notice that the free energy is always negative, becoming more negative with increasing temperature, due to the increasingly significant role of entropy. 

\begin{figure}
    \centering
    \subfigure[thermal energy $\mathcal{E}_{\beta}$]{\includegraphics[width=0.32\textwidth]{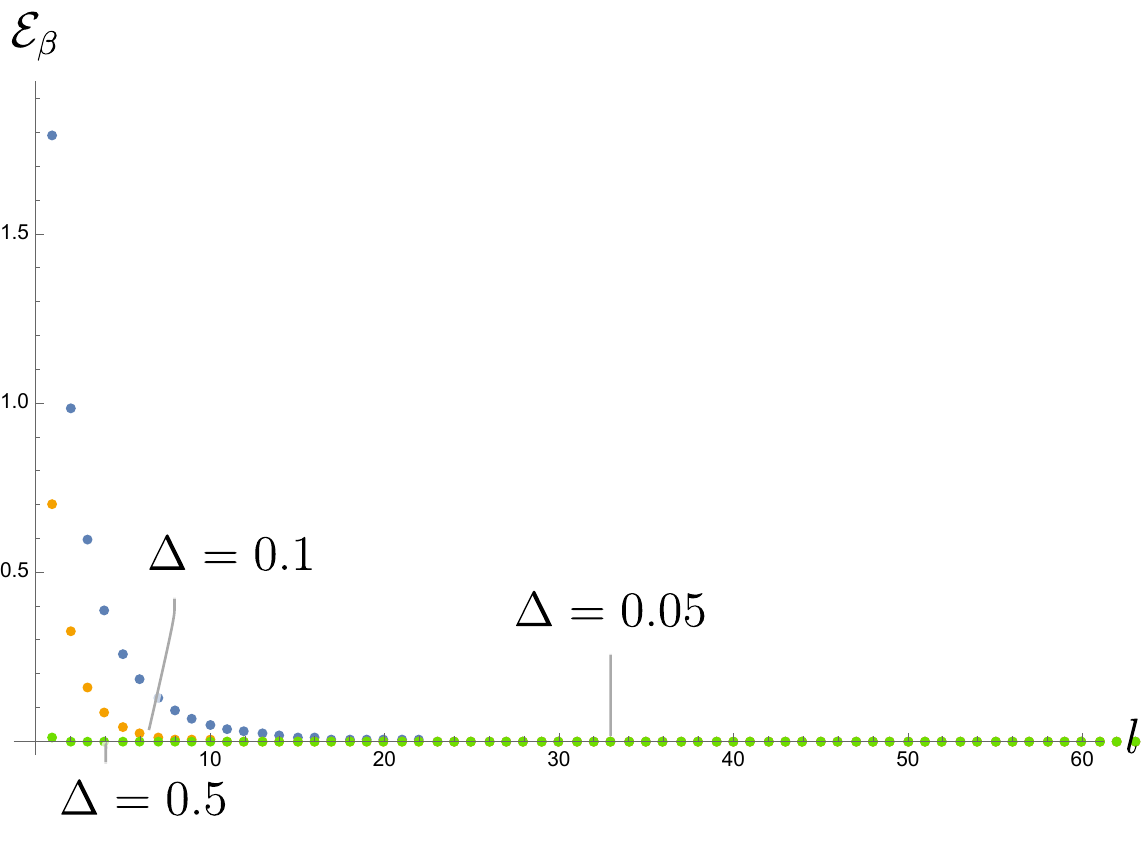}} 
    \subfigure[von Neumann entropy $\mathcal{S}$]{\includegraphics[width=0.32\textwidth]{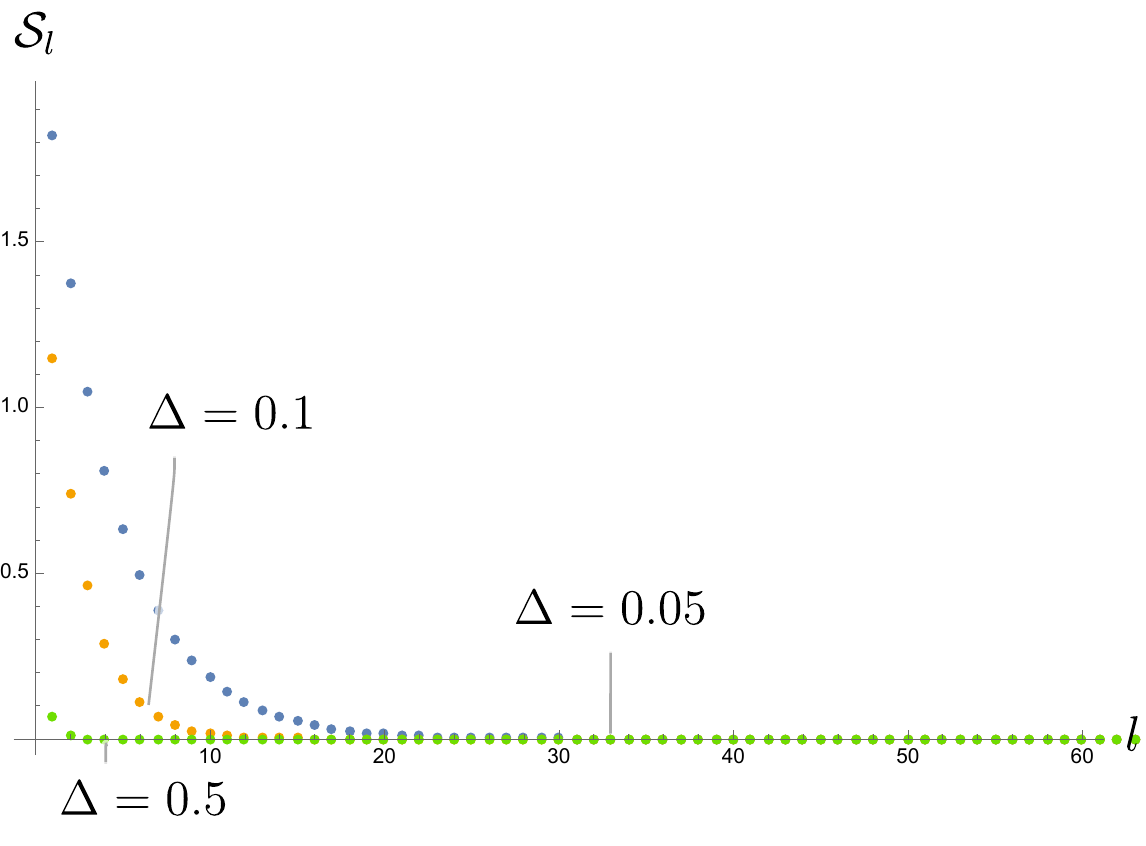}} 
    \subfigure[heat capacity $\mathcal{C}_{\textsc{v}}$]{\includegraphics[width=0.32\textwidth]{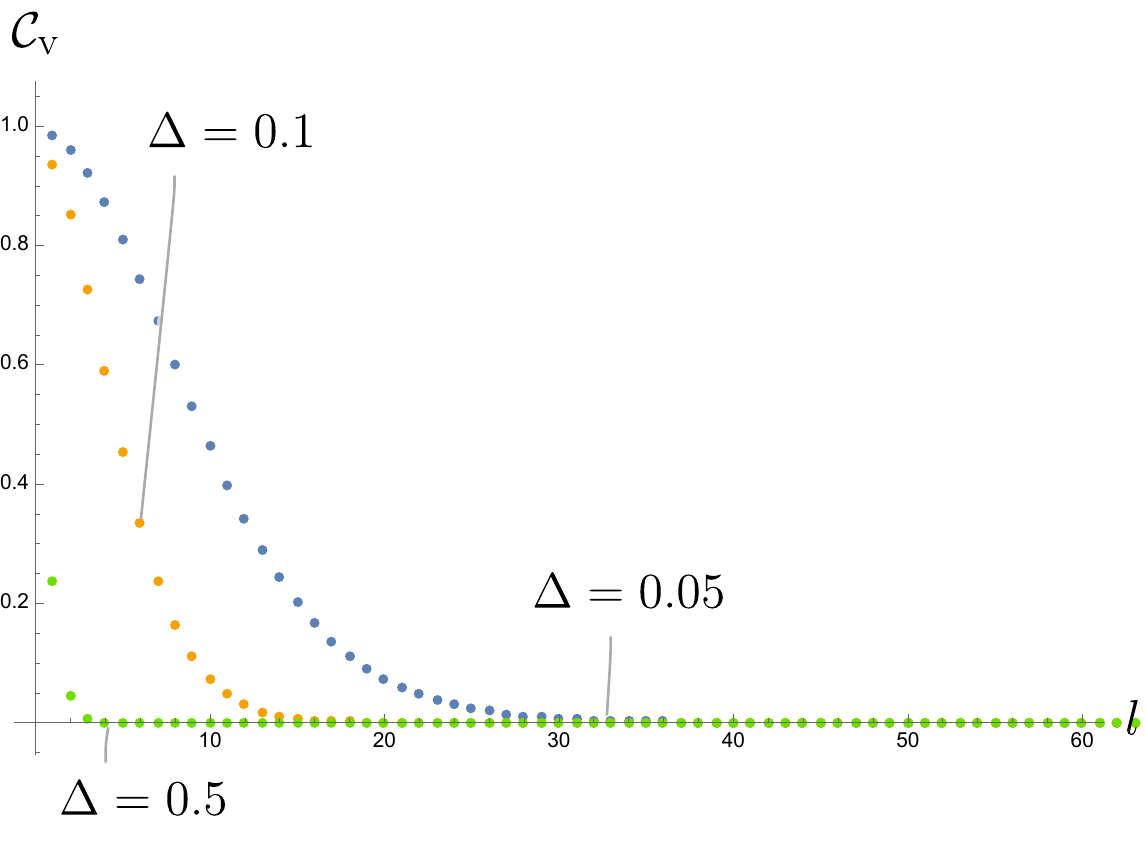}}
    \caption{Mode dependence of thermal energy, entropy and heat capacity at constant volume. In each plot, we show the curves corresponding to three choices of $\Delta$, whose inverse is proportional to the effective temperature $T_{\textrm{eff}}$ of the field after the transition is completed. These plots are consistent with the low-temperature approximations in Eqs.~\eqref{E:niue1}--\eqref{E:niue3}. The values of the parameters used here are $m=1$, $a_1=1$, $a_f=20$. For these choices, most of the modes fall within the low temperature regime  $\beta_{\textrm{eff}}\varpi_l\gg1$.}
    \label{Fi:mode}
\end{figure}

Since at late times we essentially have an effective thermal field in a constant radius $S^3$ with fixed volume $\mathcal{V}$, we can use this free energy to find the internal energy $\mathcal{E}$, entropy $\mathcal{S}$, and the other thermodynamic quantities
\begin{align}
    \mathcal{E}&=-\frac{\partial}{\partial\beta_{\textrm{eff}}}\ln\mathcal{Z}=\sum_{l=0}^{\infty}d_l^{(4)}\,\frac{\varpi_l}{2}\,\coth\frac{\beta_{\textrm{eff}}\varpi_l}{2}\,,\label{E:oebdk2}\\
    \mathcal{S}&=\beta_{\textrm{eff}}^2\biggl(\frac{\partial\mathcal{F}\hphantom{|_\textsc{ef}}}{\partial\beta_{\textrm{eff}}}\biggr)_{\mathcal{V}}= \sum_{l=0}^{\infty}d_l^{(4)}\Bigl[\frac{\beta_{\textrm{eff}}\,\varpi_l}{e^{\beta_{\textrm{eff}}\varpi_l}-1}-\ln\bigl(1-e^{-\beta_{\textrm{eff}}\varpi_l}\bigr)\Bigr]\,,\\
    \mathcal{C}_{\textrm{V}}&=-\beta_{\textrm{eff}}^2\biggl(\frac{\partial\mathcal{E}}{\partial\beta_{\textrm{eff}}}\biggr)_{\mathcal{V}}=-\sum_{l=0}^{\infty}d_l^{(4)}\frac{\beta_{\textrm{eff}}^2\varpi_l^2}{4}\,\operatorname{csch}^2\frac{\beta_{\textrm{eff}}\varpi_l}{2}\,,\label{E:oebdk4}
\end{align}
where Eq.~\eqref{E:evsdsd} can be used for the expression $\beta_{\textrm{eff}}\varpi_l$

\begin{figure}
    \centering
    \subfigure[thermal energy $\mathcal{E}_{\beta}$]{\includegraphics[width=0.32\textwidth]{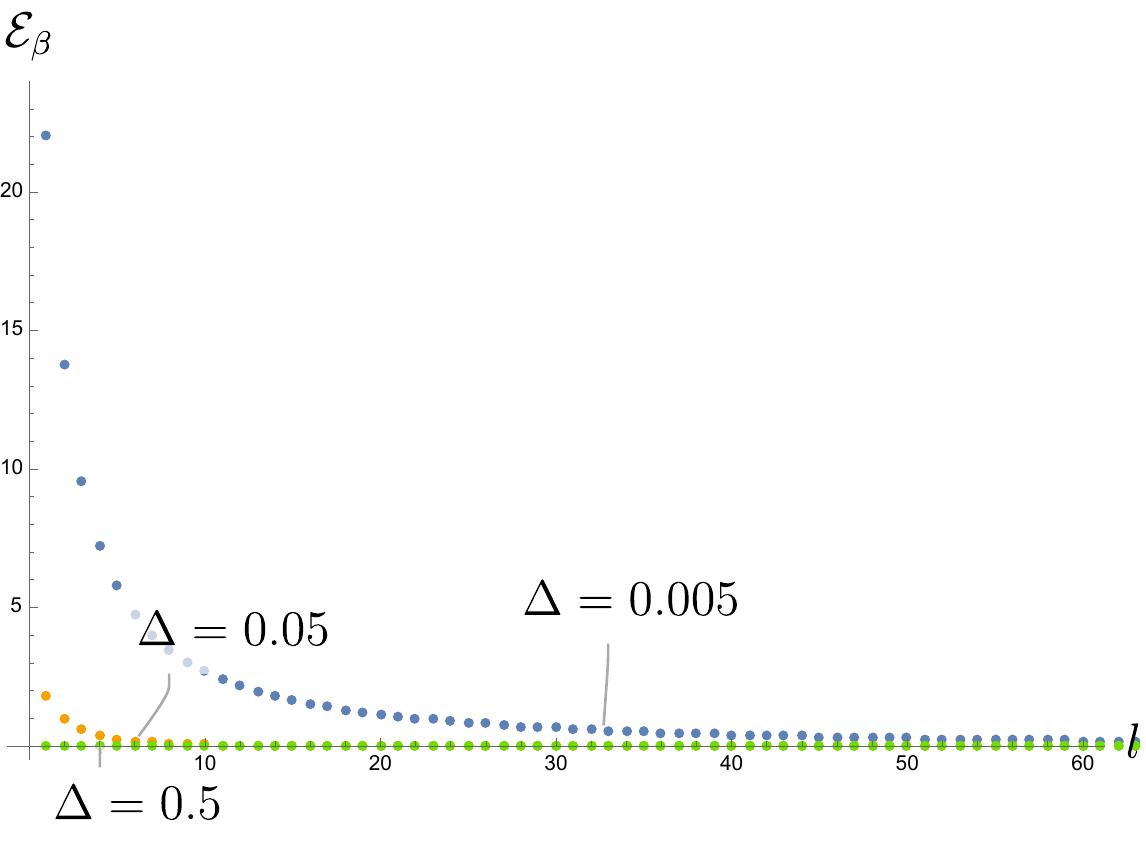}} 
    \subfigure[von Neumann entropy $\mathcal{S}$]{\includegraphics[width=0.32\textwidth]{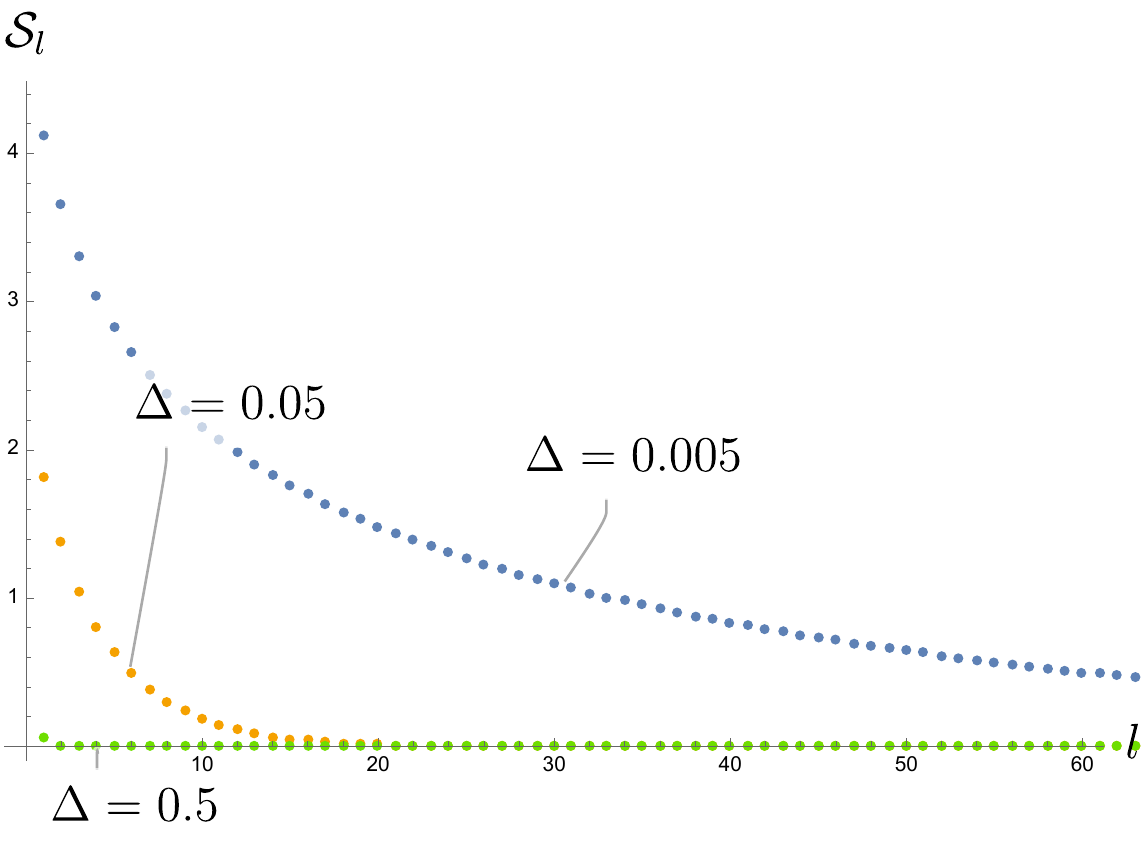}} 
    \subfigure[heat capacity $\mathcal{C}_{\textsc{v}}$]{\includegraphics[width=0.32\textwidth]{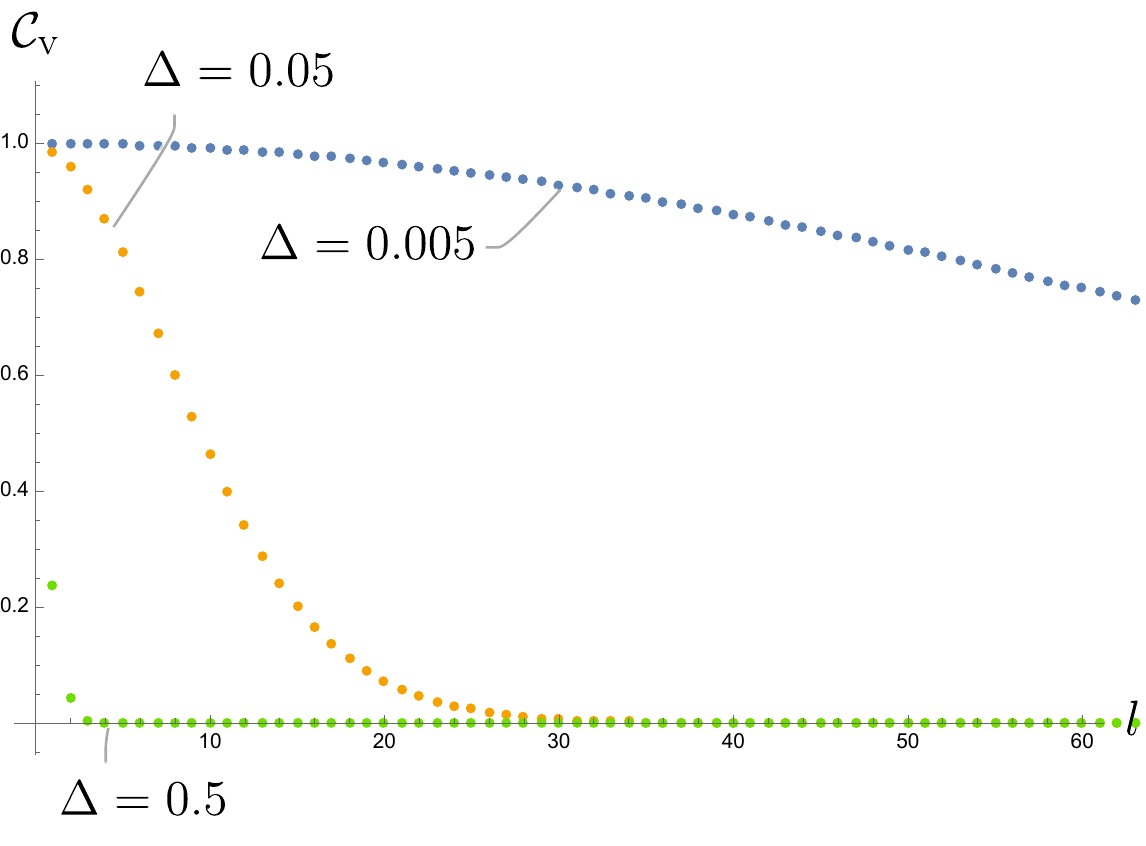}}
    \caption{Mode dependence of thermal energy, entropy and heat capacity at constant volume, similar to Fig.~\ref{Fi:mode}, except for a different $a_f=200$. For these choices, the contributions of the modes within the high temperature regime  $\beta_{\textrm{eff}}\varpi_l\ll1$ can be better seen. They are consistent with the approximations in Eqs.~\eqref{E:niue4}--\eqref{E:niue6}.}
    \label{Fi:modeHigh}
\end{figure}

The contributions of each mode in the thermodynamic quantities, Eqs.~\eqref{E:oebdk2}-\eqref{E:oebdk4} take on the familiar forms in the low- and the high-temperature limits, as follows 
\begin{itemize}
    \item low-temperature, $\beta_{\textrm{eff}}\varpi_l\gg1$:
        \begin{align}
            \mathcal{E}_l&\simeq\frac{\varpi_l}{2}+\omega_l\,e^{-\beta_{\textrm{eff}}\varpi_l}+\cdots\,,\label{E:niue1}\\
            \mathcal{S}_l&\simeq\bigl(1+\beta_{\textrm{eff}}\varpi_l+\cdots\bigr)\,e^{-\beta_{\textrm{eff}}\varpi_l}\,,\\
            \mathcal{C}_{\textrm{V}}&\simeq\bigl(\beta_{\textrm{eff}}^2\varpi_l^2+\cdots\bigr)\,e^{-\beta_{\textrm{eff}}\varpi_l}\,,\label{E:niue3}
        \end{align}
    \item high-temperature, $\beta_{\textrm{eff}}\varpi_l\ll1$:
        \begin{align}
            \mathcal{E}_l&\simeq\frac{1}{\beta_{\textrm{eff}}}+\frac{\beta_{\textrm{eff}}\varpi_l^2}{12}+\cdots\,,\label{E:niue4}\\
            \mathcal{S}_l&\simeq\ln\frac{1}{\beta_{\textrm{eff}}\varpi_l}+1+\frac{\beta_{\textrm{eff}}^2\varpi_l^2}{24}+\cdots\,,\\
            \mathcal{C}_{\textrm{V}}&\simeq1-\frac{\beta_{\textrm{eff}}^2\varpi_l^2}{12}+\cdots\,,\label{E:niue6}
        \end{align}
\end{itemize}
These are consistent with results shown in Fig.~\ref{Fi:mode} and Fig.~\ref{Fi:modeHigh}, where we have chosen the parameters such that $\pi\omega_l^{\textsc{out}}\Delta\gg1$ as discussed earlier, and generated the result for each mode according to Eqs.~\eqref{E:oebdk2} to \eqref{E:oebdk4} with $\beta_{\textrm{eff}}\varpi_l$ replaced by $2\pi\omega_l^{\textsc{in}}\Delta$.

To find the  thermodynamic quantities of the field, for simplicity we will assume $\pi ma_f\Delta\gg1$, which can be translated into
\begin{equation}
    \frac{a_f}{a_i}\frac{m}{T_{\textrm{eff}}}\gg1\,.
\end{equation}
This seems plausible for a quantum field of finite mass, in consideration of the extreme change of the scale factor under an exponential expansion. It allows the lower bound of the summation to start from $l=0$. When high frequency modes are included in the sum over all modes, the thermodynamic quantities of the field may diverge and regularization measures need be introduced, as is usually done with the calculation of the field energy\footnote{For example, the vacuum contribution to the renormalized internal energy in the out-state is given by
\begin{equation}
    \mathcal{E}_{\textsc{vac}} = \int_{ma_f}^{\infty}\!d\xi\;\frac{\xi^2}{e^{2\pi\xi}-1}\,\sqrt{\frac{\xi^2}{a_f^2}-m^2}\,.
\end{equation}}. On a closer inspection, except for the internal energy, we note that for a fixed effective temperature, the high-frequency contributions to the entropy and the heat capacity are all exponentially suppressed, so when we add up the contributions of all modes, the entropy and the heat capacity of the field  are well defined.

As for the internal energy, if we leave out the contributions of vacuum fluctuations in the out-state, and consider only the thermal energy $\mathcal{E}_{\beta}$ of created particles from the parametric amplification of the vacuum fluctuations in the in state,   then
\begin{equation}
    \mathcal{E}_{\beta}=\sum_{l=0}^{\infty}d_l^{(4)}\,\frac{\varpi_l}{e^{\beta_{\textrm{eff}}\varpi_l}-1}\,,
\end{equation}   
the result is also well defined. Since the contributions to the thermodynamic quantities from sufficiently high frequency modes are exponentially suppressed, we may truncate the infinite series to finite sums in performing numerical computations. For very small $\Delta$,  many more terms must be included in the summation in order to obtain stably convergent results.
\begin{figure}
    \centering
    \subfigure[thermal energy $\mathcal{E}_{\beta}$]{\includegraphics[width=0.32\textwidth]{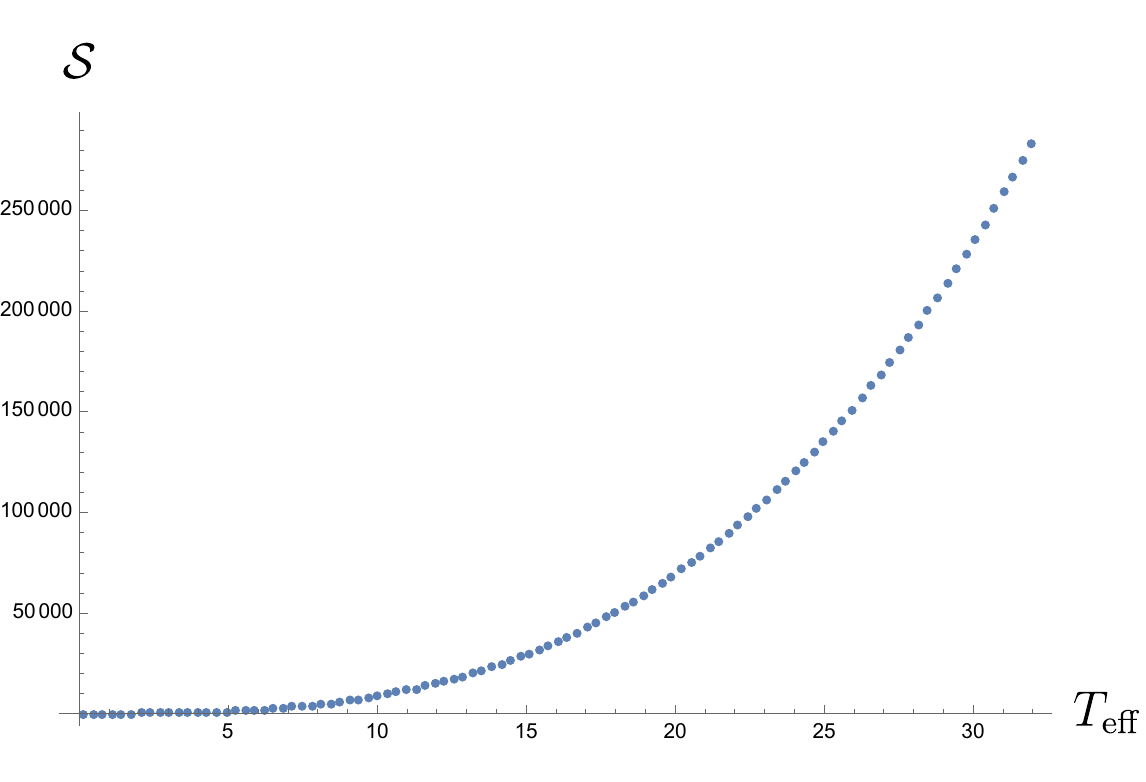}} 
    \subfigure[von Neumann entropy $\mathcal{S}$]{\includegraphics[width=0.32\textwidth]{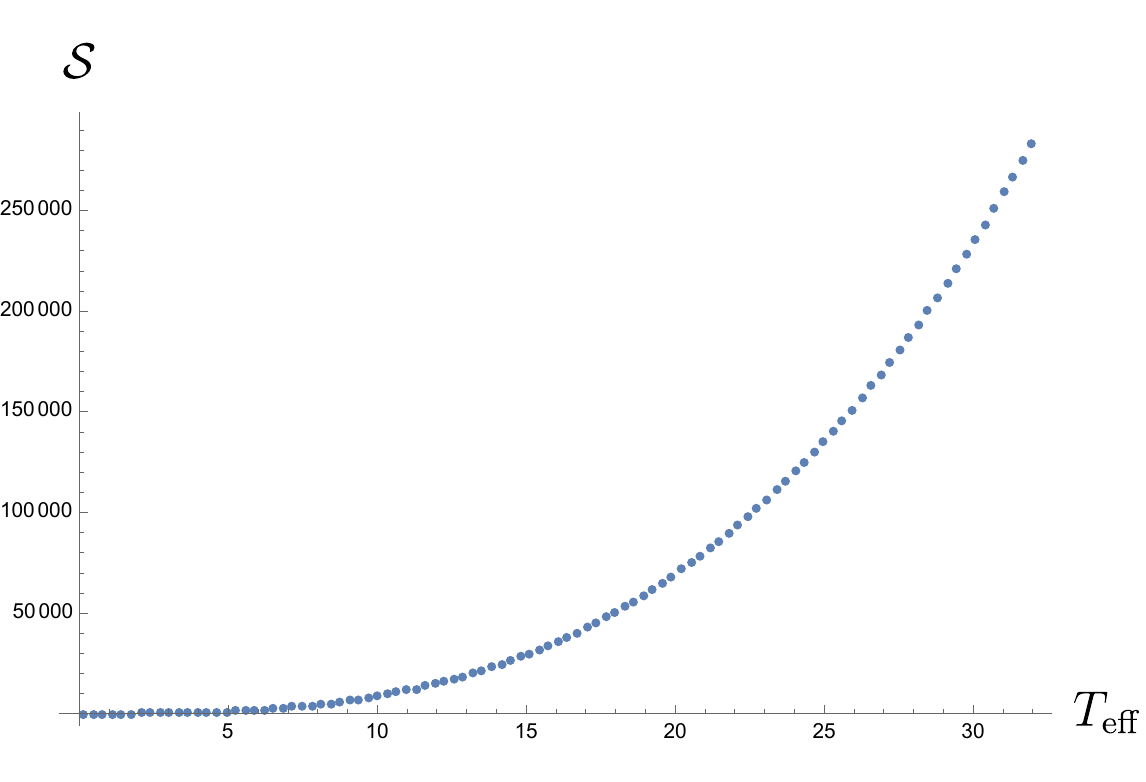}} 
    \subfigure[heat capacity $\mathcal{C}_{\textsc{v}}$]{\includegraphics[width=0.32\textwidth]{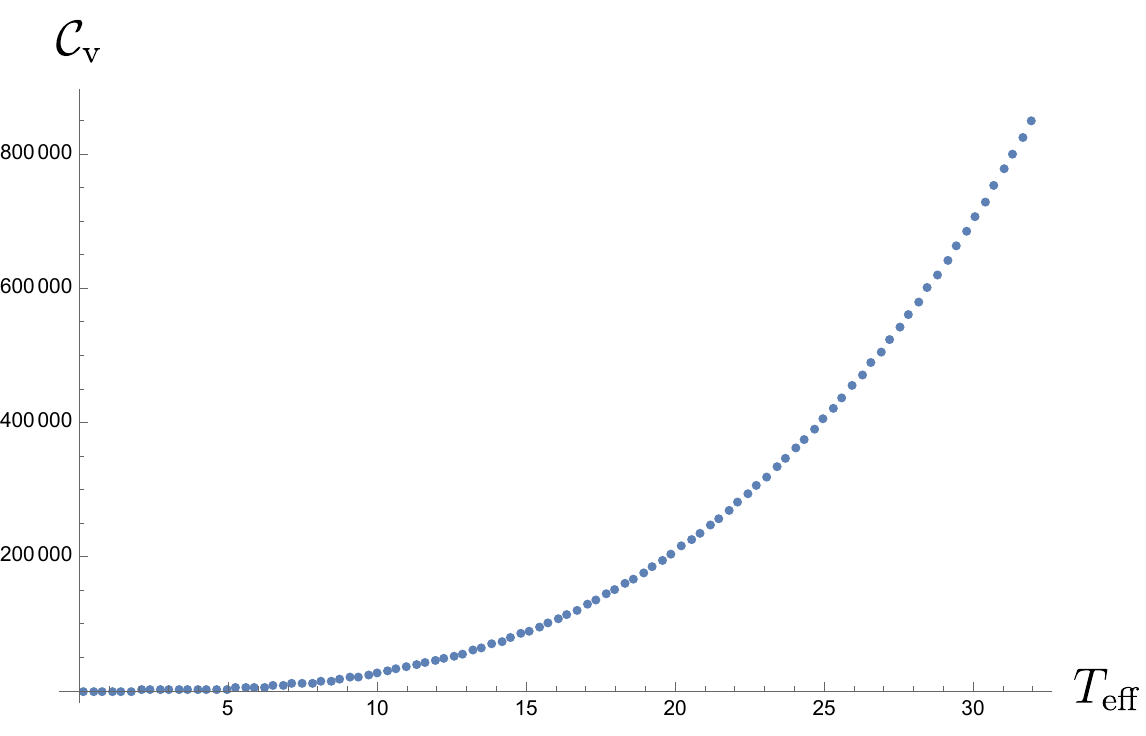}}
    \caption{Here we show $\mathcal{E}_{\beta}$, $\mathcal{S}$, and $\mathcal{C}_{\textsc{v}}$ of the whole field after we add up the contributions of all modes. We see the familiar results of $\mathcal{E}_{\beta}$ growing like $T_{\textrm{eff}}^4$ (Stefan-Boltzmann) while $\mathcal{S}$, and $\mathcal{C}_{\textsc{v}}$ grow like $T_{\textrm{eff}}^3$ for sufficiently large $T_{\textrm{eff}}$. Here we choose $a_i=1$, $a_f=200$, $m=1$,}
    \label{Fi:field}
\end{figure}

The results are shown in Fig.~\ref{Fi:field}. At late times after the transition ends, in the regime of high effective temperature, the energy scales like $T_{\textrm{eff}}^4$ and the constant volume heat capacity is proportional to $T_{\textrm{eff}}^3$. Meanwhile the entropy grows with $T_{\textrm{eff}}^3$. They behave very similar to those of bosonic particles in the relativistic limit in thermal equilibrium~\cite{KT1994} at temperature $T_{\textrm{eff}}$.

To calculate thermodynamic quantities such as the pressure, and from it the compressibility, we need variable volumes. For this purpose we consider a different scenario, namely,  We fix the initial scale factor $a_i$ and the duration of transition $\Delta$, and then examine the dependence of the Helmholtz free energy $\mathcal{F}$ on the final scale factor $a_f$. Then the pressure will be given by
\begin{equation}
    \mathcal{P}=-\biggl(\frac{\partial\mathcal{F}}{\partial\mathcal{V}}\biggr)_{T}\,,
\end{equation}
and the compressibility $\kappa_{\textsc{t}}$
\begin{equation}
    \kappa_{\textsc{t}}=-\frac{1}{\mathcal{V}}\biggl(\frac{\partial\mathcal{V}}{\partial\mathcal{P}}\biggr)_{T}\,.
\end{equation}
The volume $\mathcal{V}$ of $S^3$ is $2\pi^2a^3$, so at late times after the end of transition, we formally obtain
\begin{align}\label{E:oneor}
    \mathcal{P}&=-\sum_{l=0}^{\infty}d_l^{(4)}\,\frac{\left(\dfrac{\partial\varpi_l}{\partial a_f}\right)_T}{\left(\dfrac{\partial\mathcal{V}}{\partial a_f}\right)_{T}}\,\biggl[\frac{1}{2}+\frac{1}{\beta_{\textrm{eff}}\varpi_l}\ln\bigl(1-e^{-\beta_{\textrm{eff}}\varpi_{l}}\bigr)\biggr]\notag\\
    &=\sum_{l=0}^{\infty}d_l^{(4)}\,\frac{(l+1)^2}{6\pi^2a_f^5\varpi_l}\,\biggl[\frac{1}{2}+\frac{1}{\beta_{\textrm{eff}}\varpi_l}\ln\bigl(1-e^{-\beta_{\textrm{eff}}\varpi_{l}}\bigr)\biggr]\,,
\end{align}
where the factor $\beta_{\textrm{eff}}\varpi_l=2\pi\omega_{l}^{\textsc{in}}\Delta$ is independent of $a_f$ and $\varpi=\sqrt{(l+1)^2/a_f^2+m^2}$. In fact, Eq.~\eqref{E:oneor} gives only the partial pressure due to particle creation. Here we can see for the very high modes $l\gg ma_f$, we may approximate the fractional before the square brackets by
\begin{equation}
    \frac{(l+1)^2}{6\pi^2a_f^5\varpi_l}\sim\frac{1}{3}\frac{\varpi_l}{2\pi^2a_f^3}\,,
\end{equation}
and formally obtain
\begin{equation}
    \mathcal{P}\sim\frac{\rho_{\mathcal{F}}}{3}\,,
\end{equation}
where $\rho_{\mathcal{F}}$ is the free-energy density $\mathcal{F}/\mathcal{V}$.

\begin{figure}
    \centering
    \subfigure[fixed $a_f$]{\includegraphics[width=0.45\textwidth]{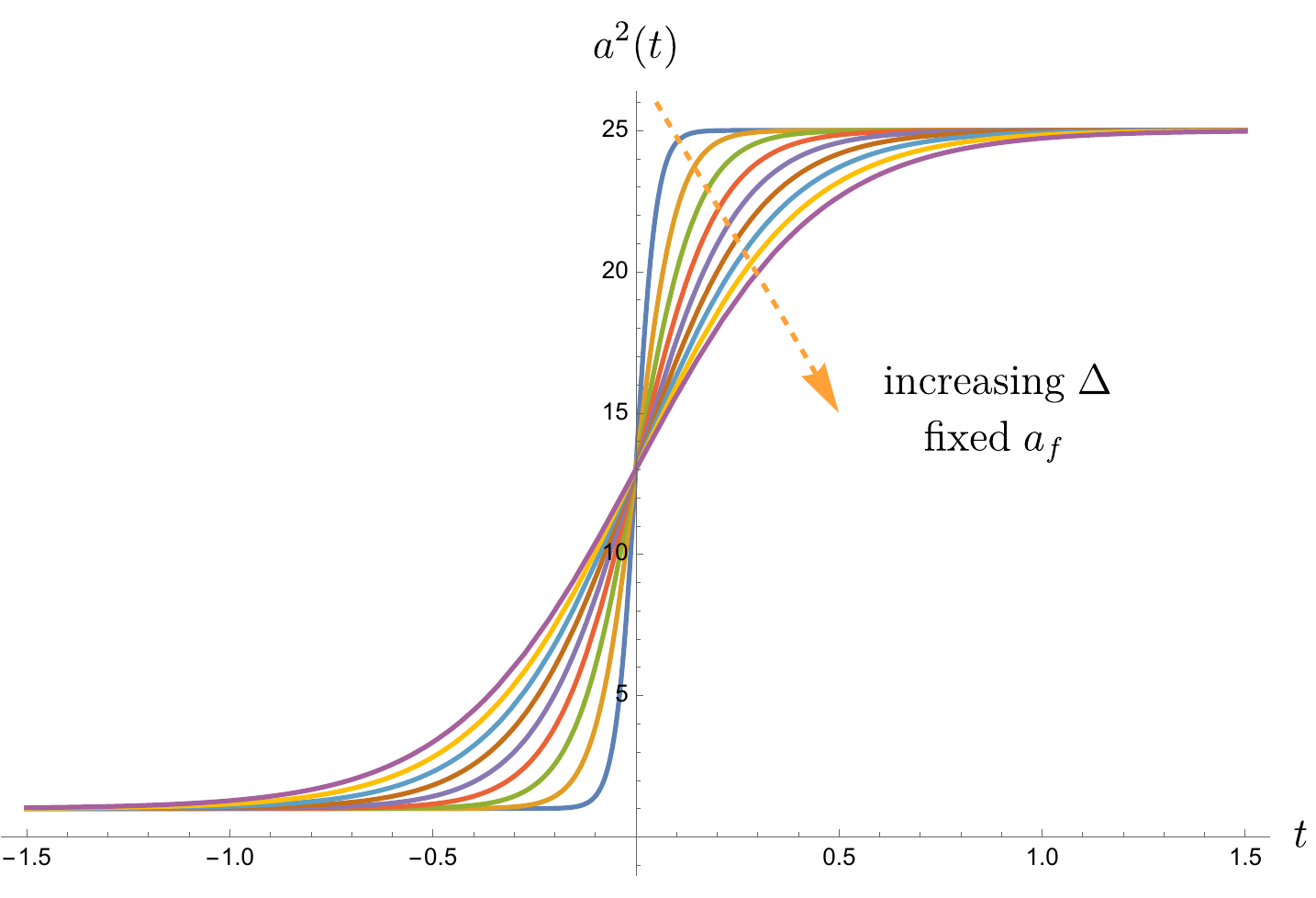}} 
    \hspace{0.05\textwidth}
    \subfigure[fixed $\Delta$]{\includegraphics[width=0.45\textwidth]{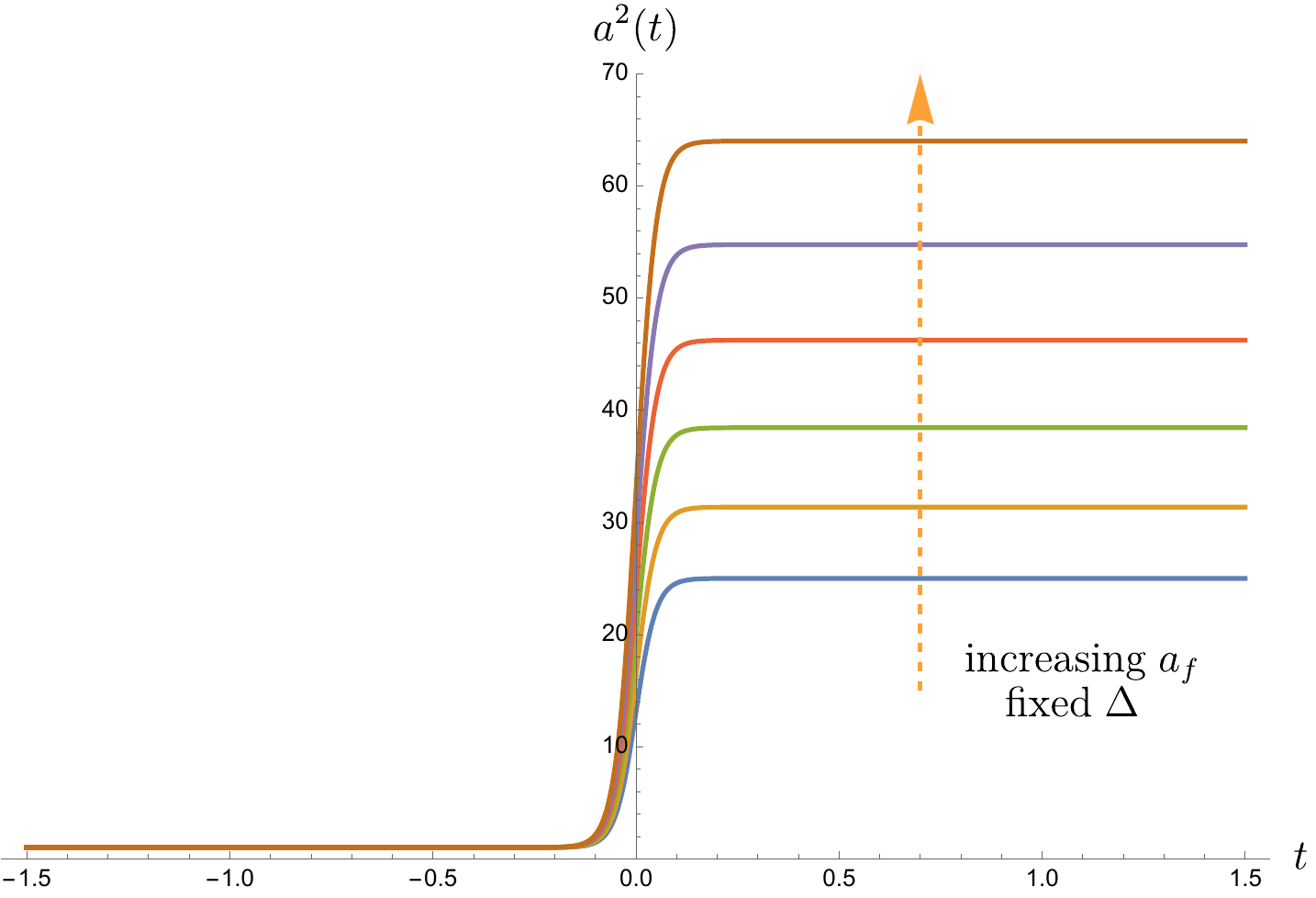}}
    \caption{Shown here are the behaviors of $a^2(t)$ for several scenarios we have considered:  In (a),  the final scale factor $a_f$ is fixed while varying $\Delta$, so that we can compare the results for different effective temperatures. In (b) the rise time $\Delta$ is fixed, while  $a_f$ takes on different values. This corresponds to keeping the temperature constant while varying the volume.}
    \label{Fi:scale}
\end{figure}
Formally, the redshift in the physical frequency due to the expansion of the spacetime,  for a fixed $\Delta$, will decrease the energy of each mode $l$, as shown in Eq.~\eqref{E:oebdk2}. There, since the factor $\beta_{\textrm{eff}}\varpi_l$ is independent of $a_f$, the mean particle number $\dfrac{1}{e^{\beta_{\textrm{eff}}\varpi_l}-1}$ of each mode and the entropy are independent of $a_f$. However,  to see the full behavior of the pressure, we  need to include  the zero-point contributions of all modes of the whole field, and deal with the divergences of  both the energy and the pressure. (See \cite{XHH1}).

If we are interested in only the thermal contribution  $\mathcal{F}_{\beta}$ of the free energy, and  $\mathcal{P}_{\beta}$ of the pressure, then by subtracting off the zero-point contributions in \eqref{E:oneor}, we find $\mathcal{P}_{\beta}$ is always negative due to the effect of redshift. This is the contribution to the isothermal compressibility from particles created via the parametric amplification of the vacuum fluctuations in the in-vacuum by the expansion of the spacetime. The result is rather counter-intuitive against our understanding of photon gas in thermal equilibrium, and against the similarities between the earlier results and the counterparts of the bosonic particles. On a closer comparison, we note that the finite-temperature part of the free energy of the photon gas is proportional to the volume  {with a negative proportionality factor}, so its $\partial\mathcal{F}_{\beta}/\partial\mathcal{V}<0$. In contrast, here due to the redshift in the frequency, we have $\partial\mathcal{F}_{\beta}/\partial\mathcal{V}>0$ instead. That makes the pressure $\mathcal{P}_{\beta}$ of the photon gas positive but renders $\mathcal{P}_{\beta}$ of the massive field negative with increasing volume.

\begin{figure}
    \centering
    \subfigure[fixed $a_f$]{\includegraphics[width=0.45\textwidth]{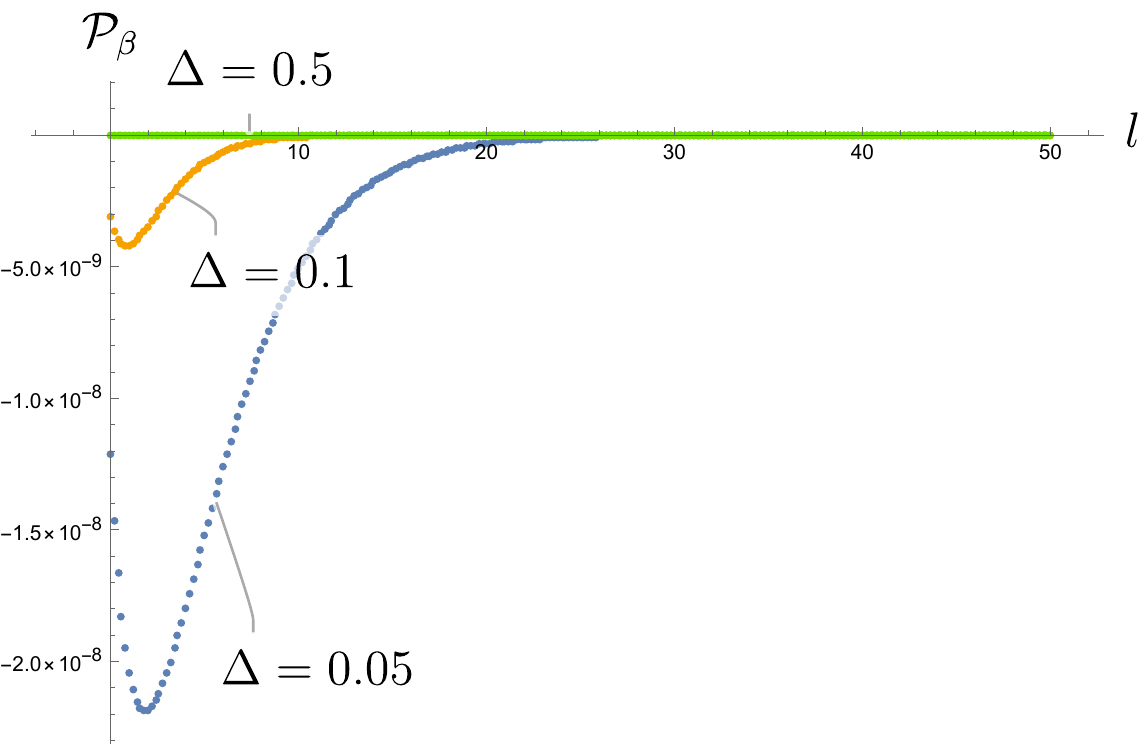}} 
    \hspace{0.05\textwidth}
    \subfigure[fixed $l$]{\includegraphics[width=0.45\textwidth]{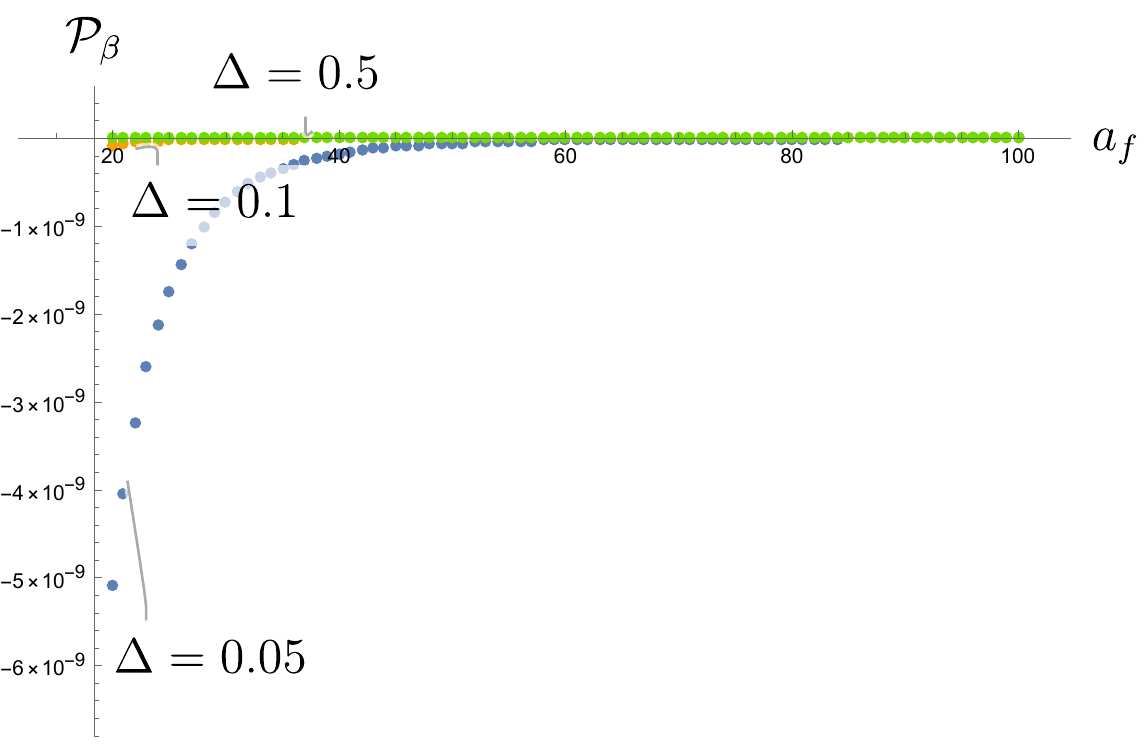}}
    \caption{(a) We show the finite-temperature contribution of the partial pressure of each mode. We choose three different effective temperatures but fix the final scale factor $a_f$. Only the modes with physical frequencies closely matching the effective temperature would be effectively excited. In (b) the partial pressure of the mode drops quickly with the spatial volume because the mean particle number is fixed by $\Delta$.}
    \label{Fi:pressure}
\end{figure}
With this understanding, now we show in Fig.~\ref{Fi:pressure}, the mode dependence of the pressure for a fixed transition duration $\Delta$ and final scale factor $a_f$ in (a) and the dependence of the pressure of a chosen mode on the final scale factor for a fixed transition interval in (b). A smaller value of $\Delta$ represents a higher value of the effective temperature, so in (a) the curve showing the dependence of the pressure on the mode has a more prominent feature for smaller $\Delta$, but drops rapidly with a greater value of $l$ because it is more difficult to excite the higher modes. In (b) we choose the mode $l=10$, and see that the pressure, although negative, falls off to zero very fast with $a_f$, Moreover, for larger values of $\Delta$ the pressure becomes extremely small. The former can be understood from the {key observation} that the mean number of particles produced $\dfrac{1}{e^{\beta_{\textrm{eff}}\varpi_{l}}-1}$ due to expansion depends only on $\Delta$ for a fixed $a_i$, and is independent of $a_f$, as seen in Eq.~\eqref{E:evsdsd}. Thus as the volume expands, the particle density is quickly diluted, leading to a decreasing pressure for each mode.

To compute the compressibility $\kappa_{\textsc{t}}$, we follow the same path to first calculate $(\partial\mathcal{P}/\partial a_f)_T$ at late times
\begin{align}\label{E:fkgbf}
   \biggl(\frac{\partial\mathcal{P}}{\partial a_f}\biggr)_T&=\sum_{l=0}^{\infty}d_l^{(4)}\,\frac{(l+1)^2}{6\pi^2}\biggl(\frac{\partial}{\partial a_f}\frac{1}{a_f^5\varpi_l}\biggr)_T\,\biggl[\frac{1}{2}+\frac{1}{\beta_{\textrm{eff}}\varpi_l}\ln\bigl(1-e^{-\beta_{\textrm{eff}}\varpi_{l}}\bigr)\biggr]\notag\\
   &=-\sum_{l=0}^{\infty}d_l^{(4)}\,\frac{(l+1)^2(4\varpi_l^2+m^2)}{6\pi^2a_f^6\varpi_l^3}\,\biggl[\frac{1}{2}+\frac{1}{\beta_{\textrm{eff}}\varpi_l}\ln\bigl(1-e^{-\beta_{\textrm{eff}}\varpi_{l}}\bigr)\biggr]\,.
\end{align}
Then we calculate the factor
\begin{equation}
    \frac{1}{\mathcal{V}}\biggl(\frac{\partial\mathcal{V}}{\partial a_f}\biggr)_{T}=\frac{3}{a_f}\,,
\end{equation}
finally arriving at {the inverse of the isothermal compressibility, that is, isothermal bulk modulus}
\begin{align}\label{E:neoir}
    \kappa_{\textsc{t}}^{-1}=\sum_{l=0}^{\infty}d_l^{(4)}\,\frac{(l+1)^2(4\varpi_l^2+m^2)}{18\pi^2a_f^5\varpi_l^3}\,\biggl[\frac{1}{2}+\frac{1}{\beta_{\textrm{eff}}\varpi_l}\ln\bigl(1-e^{-\beta_{\textrm{eff}}\varpi_{l}}\bigr)\biggr]\,.
\end{align}
{Let us first examine the expressions inside the square brackets in the low and the high effective temperature limits. The first term represents the zero-point contribution, while the second term accounts for particle creation resulting from the parametric amplification of the in-vacuum by the expansion of the spacetime. We find
\begin{equation}\label{E:btetw}
    \frac{1}{\beta_{\textrm{eff}}\varpi_l}\ln\bigl(1-e^{-\beta_{\textrm{eff}}\varpi_{l}}\bigr)=\begin{cases}
            \dfrac{\ln\beta_{\textrm{eff}}\varpi_{l}}{\beta_{\textrm{eff}}\varpi_{l}}\,,&\beta_{\textrm{eff}}\varpi_{l}\ll1\,,\;\text{high temperature regime}\,,\vspace{8pt}\\
            \dfrac{e^{-\beta_{\textrm{eff}}\varpi_{l}}}{\beta_{\textrm{eff}}\varpi_{l}}\,,&\beta_{\textrm{eff}}\varpi_{l}\gg1\,,\;\text{low temperature regime}\,.
        \end{cases}
\end{equation}
Since the factor $\ln z/z$ gives a very large negative number as $z\to 0$, in the high temperature regime, the vacuum contribution in $\kappa_{\textsc{t}}^{-1}$ can be safely ignored for modes whose values of $l$ are not too high. {However, with increasing $l$, the modes behave more like that in the low temperature regime since $\beta_{\textrm{eff}}\,\varpi_l=2\pi\omega_l^{\textsc{in}}\Delta$. The summation in \eqref{E:neoir} eventually leads to divergences for a fixed $a_f$, and renormalization procedures need be introduced. From \eqref{E:oneor}, the renormalized zero-point contribution of the pressure is given by~\cite{BordagBook}
\begin{equation}\label{E:wworo}
    \mathcal{P}_{\textsc{vac}}=\frac{1}{6\pi^2a_f^5}\int_{ma_f}^{\infty}\!dx\;\frac{x^4}{\bigl(e^{2\pi x}-1\bigr)\sqrt{x^2/a_f^2-m^2}}
\end{equation}
by the Abel–Plana formula. If we suppose $ma_f\gg1$, Eq.~\eqref{E:wworo} is approximately given by~\cite{Mamayev76}
\begin{equation}
    \mathcal{P}_{\textsc{vac}}\simeq\frac{(ma_f)^\frac{7}{2}}{12\pi^2a_f^4}\,e^{-2\pi ma_f}\,.
\end{equation}
This is a very tiny positive quantity. The corresponding term in \eqref{E:fkgbf} is
\begin{equation}
    \biggl(\frac{\partial\mathcal{P}_{\textsc{vac}}}{\partial a_f}\biggr)_T=-\frac{1}{2}\sum_{l=0}^{\infty}d_l^{(4)}\,\frac{(l+1)^2(4\varpi_l^2+m^2)}{6\pi^2a_f^6\varpi_l^3}\simeq-\frac{m^3 \sqrt{ma_f} (4 \pi ma_f+1)}{24 \pi ^2 a_f^2}\,e^{-2 \pi ma_f}\,. 
\end{equation}
Again this is exponentially small in the limit, $ma_f\gg1$. Thus we only need to consider the finite-temperature contribution in \eqref{E:neoir}, which is given by
\begin{align}\label{E:neiitoir}
    \kappa_{\textsc{t}}^{-1}\simeq\sum_{l=0}^{\infty}d_l^{(4)}\,\frac{(l+1)^2(4\varpi_l^2+m^2)}{18\pi^2a_f^5\varpi_l^3}\times\frac{1}{\beta_{\textrm{eff}}\varpi_l}\ln\bigl(1-e^{-\beta_{\textrm{eff}}\varpi_{l}}\bigr)\,.
\end{align}}

\begin{figure}
    \centering
    \subfigure[temperature dependence]{\includegraphics[width=0.45\textwidth]{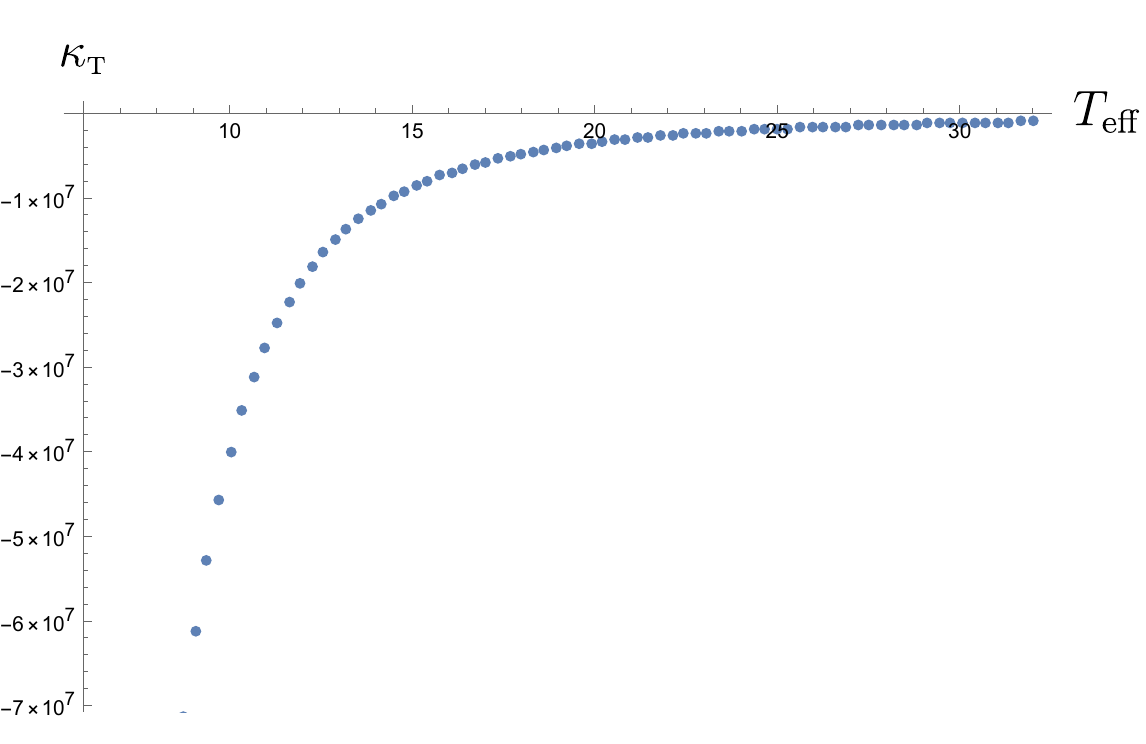}} 
    \hspace{0.05\textwidth}
    \subfigure[volume dependence]{\includegraphics[width=0.45\textwidth]{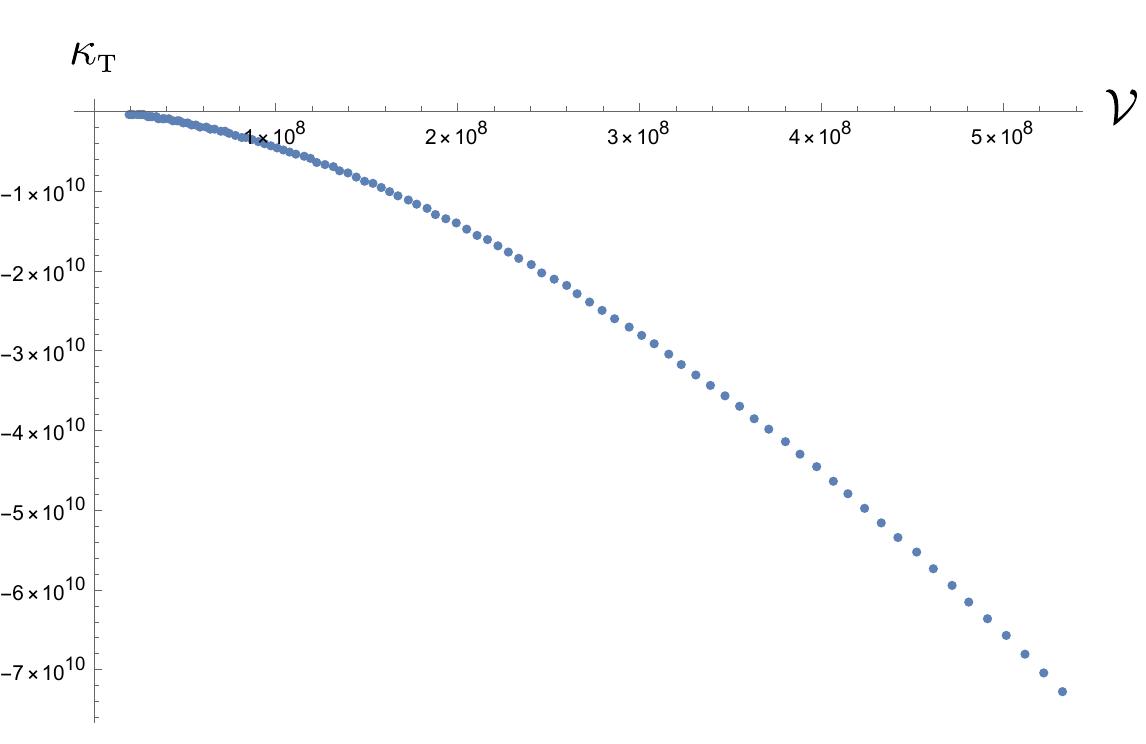}}
    \caption{Shown in (a) is the temperature dependence of the isothermal compressibility $\kappa_{\textsc{t}}$, computed by Eq.~\eqref{E:neiitoir}, leaving out the renormalized zero-point contribution. For fixed $a_f$, the space behaves `stiffer' with increasing temperature. Shown in (b) is the volume dependence of $\kappa_{\textsc{t}}$. In contrast, the space with larger volume becomes more compressible. In both plots we choose $m=1$, $a_i=1$.}
    \label{Fi:compressibility}
\end{figure}

In Fig.~\ref{Fi:compressibility}-(a), we see that the isothermal compressibility $\kappa_{\textsc{t}}$ takes on smaller negative values if the effective temperature is higher, associated with more rapid expansion {for the same $a_f$}. The physical meaning of a negative $\kappa_{\textsc{t}}$ poses some challenge\footnote{It is interesting to note that the expressions inside the square brackets in \eqref{E:neoir} can be zero when $e^{-\frac{\beta_{\textrm{eff}}\,\varpi_l}{2}}=\dfrac{\sqrt{5}-1}{2}<1$. Since $\dfrac{\beta_{\textrm{eff}}\,\varpi_l}{2}=\pi\omega_l^{\textsc{in}}\Delta$ from \eqref{E:evsdsd}, we understand that this occurs only for a specific mode for a fixed $\Delta$ and thus it will not render $\kappa_{\textsc{t}}^{-1}$ zero. In other words, the isothermal compressibility $\kappa_{\textsc{t}}$ is well defined.}. It would be totally counter-intuitive if one invokes examples of ordinary matter such as the gas dynamics of a closed system, where the volume is inversely related to the pressure such that the compressibility is positive. In the current case, the system is not closed. Expansion of the universe (or in analog gravity, an external agent which  drives the system) imparts energy into spacetime resulting in particle creation via the parametric amplification of  the vacuum fluctuations. (For a description of this process in terms of the fluctuation-dissipation relation, see \cite{CalHu87,HuSin95}.) Having this in mind, it may not be too surprising to see the thermodynamic behavior of the spacetime under study: the volume of the universe does not change as fast as the increase in pressure caused by particle creation due to the rapid expansion of the spacetime or, in analog gravity, driven by an external potential. In addition, according to \eqref{E:btetw}, a  spacetime filled with massive quanta of a conformally-coupled field would appear to be `stiffer' at higher temperatures.} {In contrast, as shown in Fig.~\ref{Fi:compressibility}-(b), at the same effective temperature, that is, with a fixed $\Delta$ the space is more compressible if it reaches a larger final volume. This characteristic is attributed to the simple fact that the particle density is more diluted in a larger spatial volume. Note that it is consistent with Fig.~\ref{Fi:pressure}-(b), where the finite-temperature contribution of the partial pressure of a specific mode drops off quickly with the final scale factor $a_f$.}

\section{Brief summary and further developments}

{In this work we have treated a conformally-coupled massive scalar field in a spacetime statically bounded in the asymptotic past and future. We assume the scale factor rises exponentially fast in conformal time from one constant and smoothly transits to anther one, as exemplified by the Bernard-Duncan model,  In the limit when the product of the mode frequency of the field in the out-state and the transition time is much greater than unity, cf. Eq.~\eqref{E:rrrhr}, once evolved from the in-vacuum in the asymptotic past, the quantum field in the out-state after such an expansion behaves nicely like a massive thermal scalar field in a Minkowski spacetime. The created particle obey the Bose-Einstein distribution at a effective temperature which is inversely proportional to the transition time. Thus, thermality results from the exponential expansion (or contraction) of spacetime, and under the thermal equilibrium condition we can work out  the thermodynamics of the quantum field in the out-state. We explicitly verify this point by examining the behavior of the Helmholtz free energy, internal energy, entropy and the heat capacity for each mode of the field and then for the field as a whole. They are consistent with the well-known results in the thermal field theory. We further examine the pressure and the isothermal compressibility. The latter is a less familiar notion in this context. We find that thermal particle creation under the rapid expansion of spacetime results in a negative value in the finite-temperature contribution to the pressure. We see this as a consequence of the redshift in the physical frequency of the modes of the field. Thus it gives a  counteracting component to the total pressure. The compressibility typically takes on small negative values. This is understood as the fact that under the current scenario, the volume does not increase as fast as the pressure does when the spacetime is expanded to various sizes over a fixed transition time. Moreover spacetime  filled with massive quanta of a conformally-coupled field exhibits increased stiffness at higher temperatures.}

What we have presented here is only a model calculation, taking advantage of the simple analytic form of the scale function $ a^2 (\eta) \sim \tanh (\eta / \Delta)$ working in conformal time $\eta$.  Two extensions of more realistic significance  can readily be made:  One is to consider a massive conformal field in spatially-flat FLRW universe with an exponential expansion stage in {\it cosmic} time $t$. The other extension is to treat a massless minimally-coupled scalar field, which depicts one polarization of the graviton field, in the Poincare patch of de Sitter space, also in  cosmic time \footnote{Note the thermal  particle creation in an inflationary universe with a period of exponential expansion where the scale factor $a(t) \sim e^{Ht}$ in the flat FLRW coordinatization of de Sitter space  is different from the thermal radiance felt by an observer in a static coordinatization of the de Sitter space, the former is due to parametric amplification of vacuum fluctuations whereas the latter is due to the Gibbons-Hawking effect akin to the Hawking effect in black holes or the Unruh effect for uniformly accelerated detectors. For a broader viewpoint of de Sitter thermality and calculations of thermodynamic functions, see, e.g., \cite{Volovik}.} The calculations will be a bit more involved, but the track laid down here toward deriving the thermodynamic functions can be used just as well. The crucial observation is that one does not need to impose the somewhat artificial static out-state, because most of the particle production happens in the initial exponential expansion stage.  As long as the particle production has a thermal spectrum the procedures illustrated here can be adopted. Alternatively, an even better and conceptually clearer approach is to adopt the in-in or closed-time-path or Schwinger-Keldysh formalism, as we alluded to in the beginning. Only a well defined initial state, or unambiguous Fock state is needed and the nonequilibrium dynamics will guide the ensuing evolution of the physical observables with respect to the initial state. At least for  Gaussian systems, following the strategy outlined in~\cite{HHNEqFE}, the final state is always in the form of a squeezed state, from which one can extract the effective temperature and the coherent, squeezed parameters up to a constant. The in-in formalism does not require the spacetime to be asymptotically-stationary as it does in an in-out $S$-matrix formulation.     

Zooming out to the bigger picture, as we explained in our earlier paper \cite{XHH1} (see references cited therein) spacetime thermodynamics offers a refreshing and perhaps even correct way of viewing the nature of spacetime, namely, as an effective theory which describes the low energy, long wavelength limit of quantum gravity -- theories about spacetime's microscopic constituents.  The novel notions of quantum capacity and dynamical compressibility introduced there and the theoretical results obtained herewith associated with thermal particle production have bearings on two ends of a broad spectrum, namely, the conceptualization of emergent gravity and the observational possibilities in analog gravity, specifically, dynamical Casimir effect experiments in the laboratory. Their physical meanings need be savored more and expounded further in these contexts.

In the study of spacetime quantum thermodynamics of dynamical spacetimes with particle creation,  two processes we mentioned in the Introduction are worthy of further investigation: For one, the inclusion of stimulated particle creation.  So far we have only considered spontaneous particle creation from the vacuum, i.e., parametric amplification of   the zero-point energy due to the expansion (or contraction) of the universe.  Even the thermal field in this paper is of vacuum origin, albeit the spectrum of particles created obey a Planck distribution, thanks to a period of exponential expansion.  To be complete,  stimulated production, i.e., parametric amplification of  particles already present needs be included at every moment. This is because at any moment whatever amount of particles created spontaneously from the vacuum  at  an earlier moment will also be amplified.  This will greatly enhance (for bosons) the amount of energy produced.  In fact, we think the great enhancement factor reported in \cite{Ralf} whose authors regard as owing to finite temperature effect, is fundamentally due to stimulated particle production.  Here, one sees the theoretical difference between thermal particle production and stimulated particle production: the stimulated production mechanism is more basic and general, even though it  applies to thermal particle creation, it is not limited to thermal fields.  Thermal field refers to the state of the quantum field while stimulated production refers to the dynamical process. It is this stimulated amplification process, like stimulated emission in lasers, which results in a much stronger effect than vacuum creation, like spontaneous emission. 

The other task is the application of nonequilibrium quantum field theory methods (e.g., \cite{CalHu2008}) for a fuller treatment of quantum field processes in the early universe.  Nonequilibrium treatments do not rely on thermal notions nor require equilibrium states as pre-conditions.  In fact there is no place for the notion of temperature -- temperature is a derived concept, taking shape only after thermalization can be actuated.  Quantum field processes like particle creation in evolutionary cosmology or in dynamical Casimir effects are intrinsically time-dependent processes, whereas finite temperature theories are predicated upon the existence of a thermal equilibrium state under stationarity conditions.  It is now possible to treat the thermodynamics of quantum field processes in dynamical spacetime from a nonequilibrium quantum field theory approach because a nonequilibrium free energy density functional related to the influence action has recently been established for dynamical quantum systems interacting with a quantum field environment \cite{HHNEqFE}. The approach and results reported there are useful for the investigations in the present research program.  We hope to report on progresses along these two fronts in our future works \cite{BHH1,CHH2}.{}\\

\noindent{\bf Acknowledgment}   J.-T. Hsiang is supported by the National Science and Technology Council of Taiwan, R.O.C. under Grant No.~NSTC 112-2112-M-011-001-MY3. B.-L. Hu enjoyed the warm hospitality of Prof.~Chong-Sun Chu of the National Center for Theoretical Sciences at National Tsing Hua University, and Prof. Kin-Wang Ng of the Institute of Physics, Academia Sinica, Taiwan, R.O.C. where part of this work was done.  

\newpage

\bibliography{refs}

\end{document}